\documentclass[preprint]{elsarticle}
\graphicspath{ {./figures/} }
\usepackage{hyperref}
\usepackage{float}
\usepackage{verbatim} 
\usepackage{apalike}
\usepackage{mathptmx}
\usepackage{latexsym}
\usepackage{alltt}
\usepackage{algorithmic}
\usepackage{algorithm}
\usepackage{amsmath,amssymb,amsthm,mathtools}
\usepackage{multirow}
\usepackage{breqn}
\usepackage[dvipsnames]{xcolor}

\usepackage{subfigure}
\usepackage{verbatim} 
\usepackage{multicol}        
\usepackage[bottom]{footmisc}

\usepackage{makeidx}         
\usepackage{graphicx}        
\usepackage{multicol}        
\usepackage[bottom]{footmisc}
\usepackage{algorithmic}
\usepackage{algorithm}
\usepackage{amsmath,amssymb}
\DeclareMathOperator*{\argmax}{argmax}
\usepackage{multirow}
\usepackage{subfigure}
\usepackage{amsthm}
\usepackage{eqnarray}
\usepackage[misc,geometry]{ifsym} 
\usepackage{calrsfs}
\DeclareMathAlphabet{\pazocal}{OMS}{zplm}{m}{n}
\restylefloat{figure}
\restylefloat{table}

\journal{Expert Systems with Applications}

\biboptions{authoryear}

\begin{document}
\begin{frontmatter}

\begin{titlepage}
\begin{center}
\vspace*{1cm}

\textbf{ \large Learning Geometric Combinatorial Optimization Problems using  \\ Self-attention and Domain Knowledge}

\vspace{1.5cm}

Jaeseung Lee$^{a}$ (hunni10@inu.ac.kr), Woojin Choi$^a$ (202027009@inu.ac.kr), \\ Jibum Kim$^a$ (jibumkim@inu.ac.kr) \\

\hspace{10pt}

\begin{flushleft}
\small  
$^a$ Department of Computer Science and Engineering, Incheon National University, Incheon, 406-772, South Korea \\

\vspace{1cm}
\textbf{Corresponding Author:} \\
Jibum Kim \\
Department of Computer Science and Engineering, Incheon National University, Incheon, 406-772, South Korea\\
Tel: +82 (32) 835-8655 \\
Email: jibumkim@inu.ac.kr

\end{flushleft}        
\end{center}
\end{titlepage}

\title{Learning Geometric Combinatorial Optimization Problems using Self-attention and Domain Knowledge}

\author[label1]{Jaeseung Lee}
\ead{hunni10@inu.ac.kr}

\author[label1]{Woojin Choi}
\ead{202027009@inu.ac.kr}

\author[label1]{Jibum Kim \corref{cor1}}
\ead{jibumkim@inu.ac.kr}

\cortext[cor1]{Corresponding author.}
\address[label1]{Department of Computer Science and Engineering, Incheon National University, Incheon, \\ 22012, South Korea}

\begin{abstract}
Combinatorial optimization problems (COPs) are an important research topic in various fields. In recent times, there have been many attempts to solve COPs using deep learning-based approaches. We propose a novel neural network model that solves COPs involving geometry based on self-attention and a new attention mechanism. The proposed model is designed such that the model efficiently learns point-to-point relationships in COPs involving geometry using self-attention in the encoder. We propose efficient input and output sequence ordering methods that reduce ambiguities such that the model learns the sequences more regularly and effectively. Geometric COPs involve geometric requirements that need to be satisfied. In the decoder, a new masking scheme using domain knowledge is proposed to provide a high penalty when the geometric requirement of the problem is not satisfied. The proposed neural net is a flexible framework that can be applied to various COPs involving geometry. We conduct experiments to demonstrate the effectiveness of the proposed model for three COPs involving geometry: Delaunay triangulation, convex hull, and the planar Traveling Salesman problem. Our experimental results show that the proposed model exhibits competitive performance in finding approximate solutions for solving these problems.
\end{abstract}


\end{frontmatter}

\section{Introduction}
Combinatorial optimization problems (COPs), in which an optimal subset is found from amongst a finite set of objects, is an important research topic with various applications. The goal of COPs is to find the optimal object that minimizes or maximizes some objective functions. Several well-known examples of COPs include the traveling salesman problem (TSP), knapsack problem, minimum spanning tree, and vehicle routing problem. Many COPs are NP problems, and heuristic algorithms are often used to find approximate solutions when solving them. In recent times, various efforts have been made to solve COPs using state-of-the-art artificial intelligence techniques~\cite{12,13,14,44,45,46,47,48}. Supervised learning and reinforcement learning have been used to solve these problems, but their effectiveness varies depending on the problem. Supervised learning is effective when an optimal solution is available and sufficient data are available for training. However, in some cases, it is not easy to obtain a high-quality labeled dataset. The supervised learning-based approach is complex, and scalability is poor when the problem is large or NP. For such problems, many reinforcement-learning-based methods have been proposed. 

Among various COPs, we consider finding approximate solutions for COPs involving geometry, where the positional relationship among points is important, using a neural net. In particular, we focus on finding approximate solutions for the Delaunay triangulation (DT) problem. DT has various applications in diverse fields. It is often used to generate triangular meshes when solving partial differential equations using the finite element method or finite volume method for a given geometric domain.

Research on generating DT using deep learning has not been studied extensively, owing to several reasons. First, when the solution of DT is expressed as an input/output sequence, the input/output length varies depending on the problem size. However, conventional seq2seq-based models are not effective for this variable-length dictionary. Second, the mesh data structure is not a regular data structure, such as an image, but is an irregular data structure, making learning difficult for the model. Applying unmodified deep learning methods, such as the convolutional neural network, is challenging~\cite{10}. Third, it is possible to predict the coordinates of a point when creating a mesh using a deep learning method, such as the RNN, but in this case, there is no limit to the feature space, which causes blurring~\cite{2}. Finally, the elements constituting the triangular mesh consist of three points, which must be learned as a group, and it is difficult for the model to learn the relationships and dependencies between points in distant locations. In particular, as the number of points forming a point set increases, the sequence length becomes longer, making it challenging to learn the dependencies between points in distant locations in the sequence.

A representative supervised learning-based deep learning model for solving COPs involving geometry is the pointer network (Ptr-Net)~\cite{2}. The conventional sequence-to-sequence (seq2seq) model computes conditional probabilities from a fixed dictionary. Therefore, it faces limitations in solving COPs in which the size of the output dictionary changes depending on the input size. Ptr-Net overcomes this issue by adapting the attention mechanism to create pointers to elements in the input sequence~\cite{21,29}. Vinyals et al. applied Ptr-Net to find approximate solutions to several COPs with variable-sized inputs and output dictionaries, but it has limitations for practical use because of its low accuracy~\cite{2}. Recently, a transformer model has been proposed to solve the long-term dependency problem of the recurrent neural network (RNN) model~\cite{3}. The transformer model consists of self-attention, multi-head attention, and positional encoding. It has the advantage that it does not involve recurrence. The transformer model is widely used in natural language processing, voice recognition, and image recognition~\cite{4, 5, 6}. The concept of self-attention used in the transformer model indicates that attention is carried out toward itself and can be used in encoder and decoder models. The transformer model learns by considering the relationships between all pairs in the input/output sequence.

In this paper, we propose a novel neural network model based on self-attention and domain knowledge that can find approximate solutions of COPs involving geometry. The sequence order of input and output has a noticeable effect on the performance of the seq2seq-based models. However, the effect of applying input/output sequence ordering is not well studied. We propose an efficient ordering method for input and output sequences that can substantially improve the learning performance of the model for various COPs involving geometry. Second, we apply self-attention in the encoder to better learn the point-to-point relationship in the input sequence. For COPs involving geometry, self-attention in the encoder is necessary because the model learns the topological and geometric relationship between distant points using self-attention better and faster. In particular, when the number of points increases and the length of the input sequence length increases, it is more difficult for the model to learn such an input sequence without self-attention in the encoder. Finally, we develop a new masking scheme that can augment the existing attention mechanism using domain knowledge.  

We investigate whether the proposed neural net model learns other COPs involving geometry. First, we consider finding the approximate solutions of the convex hull problem using the proposed neural net model. The convex hull of a point set is the smallest convex object that includes the point set. We consider the convex hull problem because DT and the convex hull problem are closely related problems, in that the DT of a given point set is a triangulation of the convex hull. It is important for the model to learn the positional relationship among points for the convex hull problem. Finally, we test whether the proposed neural net is able to learn and produce competitive approximate solutions for TSP. Specifically, we consider finding approximate solutions of 2D planar symmetric TSP, where the distance between two cities is the same in each opposite direction. Its goal is to find the shortest possible route that visits each city exactly once and then returns to the original city. It is a classical NP-hard problem even in the 2D Euclidean case. Our main contributions are summarized as follows:

\begin{itemize}
\item We propose a novel neural network model that learns COPs involving geometry based on self-attention and domain knowledge. The proposed model is the first that applies a self-attention mechanism for learning DT in the encoder. We also propose an efficient input/output sequence ordering method to learn the sequence more regularly and effectively.
\item We propose a new masking scheme using domain knowledge that can be applied to general COPs involving geometry. We also develop several evaluation metrics that measure model performance.
\item The proposed model is a flexible framework that can be applied to various COPs involving geometry. We show that the proposed model learns and produces a competitive solution for NP problems such as TSP. 
\end{itemize}

The rest of the article is organized as follows. Section~\ref{sec:Related} describes related works. In Section~\ref{sec:Background}, we introduce existing neural net models including Ptr-Net. In Section~\ref{sec:Model}, we describe the proposed neural net model based on self-attention and a new attention mechanism. In Section~\ref{sec:Experiments}, the experiments, dataset, and metrics are explained in detail. Section~\ref{sec:Results} shows experimental results for three COPs involving geometry: DT, convex hull, and TSP. We discuss the limitations of our model in Section~\ref{sec:Discussion}. We conclude our paper in Section~\ref{sec:Conclusion}.

\begin{figure}[t]
\centering
\hspace{1cm}
\includegraphics[width=9cm]{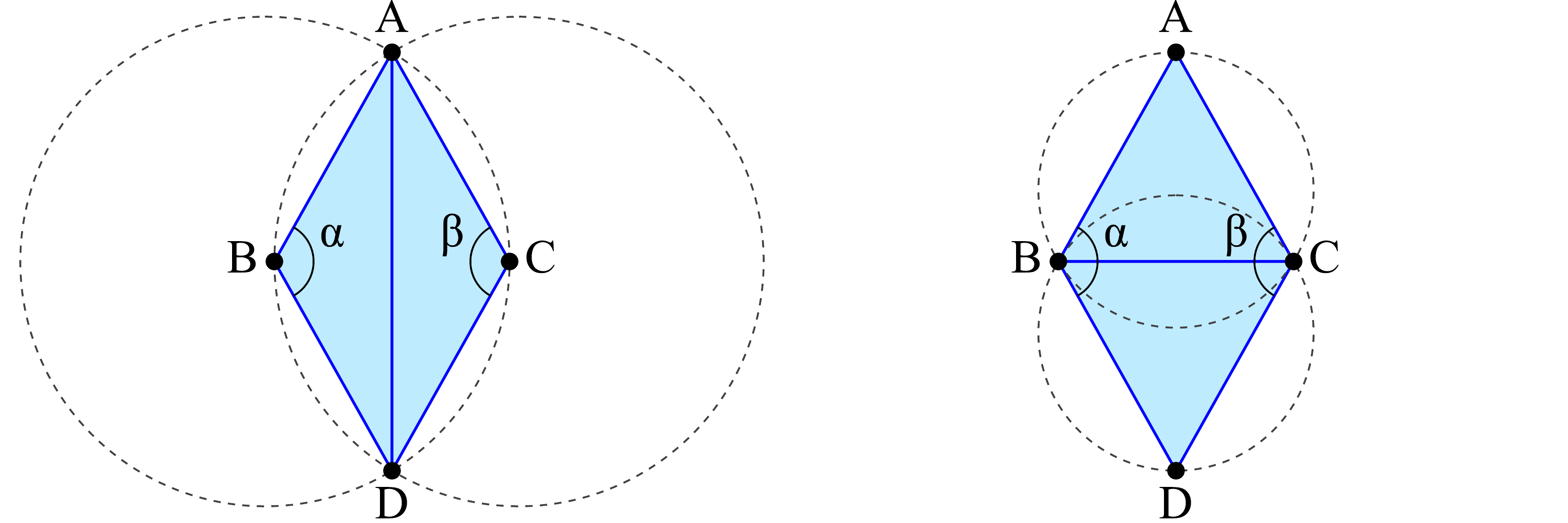}
\caption{Triangulation that (left) does not meet the Delaunay condition and (right) meets the Delaunay condition after performing an edge flip.}
\label{fig:fig_2}
\end{figure}

\section{Related Work}
\label{sec:Related}
{\noindent \bf  {Delaunay Triangulation.} } Many algorithms have been proposed for generating DT with the algorithm complexity of  $O\left ( n\mathrm{log}(n) \right )$ or $O\left ( n^2 \right )$ but most algorithms are non-trivial algorithms~\cite{3,7,8,9,30}. The CGAL package can be used for generating DT~\cite{36}. One of the most popular DT algorithms is the flip algorithm, which constructs any triangulation from the points and then repeatedly flips the edges until every triangle meets the Delaunay condition. When two triangles ABC and ABD share an edge AB, if the sum of the $\alpha$ and $\beta$ angles is less than or equal to $180^{\circ}$, the Delaunay condition is satisfied. Figure~\ref{fig:fig_2} displays two triangulations, where one triangulation (left) does not meet the Delaunay condition and the other triangulation (right) meets the Delaunay condition after performing an edge flip (swap).

To the best of our knowledge, two deep learning-based approaches can generate DT from a given input point set with limited performance. Both methods are based on supervised learning and use the encoder-decoder model. These methods  automatically predict DT for a given point set after the model sufficiently learns DT. In~\cite{2}, the authors used the Ptr-Net for generating DT. It obtains 80.7\% accuracy and 93.0\% triangle coverage when the number of points ($m$) is 5. However, when $m$ = 10, the accuracy dramatically decreases to 22.6\%, and the triangle coverage decreases to 81.3\%. In addition, when $m$ = 50, it does not produce any precisely correct triangulation, and the triangle coverage is also significantly reduced to 52.8\%.

In~\cite{11}, the authors proposed the multi-Ptr-Net (M-Ptr-Net), a variant of the Ptr-Net, which uses multiple ``pointers'' to learn by combining the vertices that constitute a triangle into a group  for the DT problem. It uses multiple ``pointers'' using the multi-label classification idea~\cite{11}. It fits a loss function of the multi-label classification using the sigmoid function instead of the softmax function. They reported that when $m$ = 5, the accuracy improves by 0.45\% compared to the Ptr-Net. However, two existing deep learning-based approaches for generating DT still have limitations for practical use due to the low accuracy and triangle coverage. In addition, the model does not sufficiently learn point-to-point relationships forming triangles. \\

{\noindent \bf  {Convex Hull.} }
There exist several algorithms that find exact solutions for convex hull problems~\cite{33,34}. It has complexity $O\left ( n\mathrm{log}(n) \right )$. There exists only one previous work that finds approximate solutions for the convex hull problem using deep learning~\cite{2}. When $m$ (number of points)=5, it achieves 92.0\% accuracy, but the accuracy drops to 50.3\% when $m$=100. \\

{\noindent \bf  {TSP.} }
The best-known exact algorithm for TSP has a complexity of $O\left (2^{n}n^{2} \right )$, which is nearly infeasible to scale up~\cite{15}. In practice, many heuristic algorithms are used to find competitive approximate solutions. Many deep-learning-based approaches can find approximate solutions of TSP. The authors reported that when the number of cities ($m$) is small (e.g., $m$=5 or 10), the performance of the model shows almost optimal performance in terms of the tour (path) length, but it shows limited performance when the number of cities $m$=50~\cite{20}. The reinforcement learning-based approach shows better performance than supervised learning for the TSP as high-quality labeled data is not available and is costly for NP problems. The authors proposed a framework to solve COPs using neural networks and reinforcement learning~\cite{15}. The authors reported that their approach improves supervised learning-based approaches and also surpasses existing heuristic algorithms~\cite{15}. The method solves the TSP and demonstrates nearly optimal results on two-dimensional Euclidean graphs for 33 - 100 nodes (cities). 

Recently, Nowak et al. formulated TSP as a quadratic assignment problem and solves it using  graph neural network (GNN) models~\cite{41}. Joshi et al. proposed a novel supervised learning-based approach that solves 2D Euclidean TSP using Graph Convolutional Networks (GCN)~\cite{14}. The authors adapted the Transformer model to solve 2D Euclidean TSP and trained using reinforcement learning~\cite{38}. The GNN-assisted Monte Carlo Tree Search approach was employed to solve the TSP~\cite{39}.

\begin{figure}[t]
\centering
\includegraphics[width=4cm]{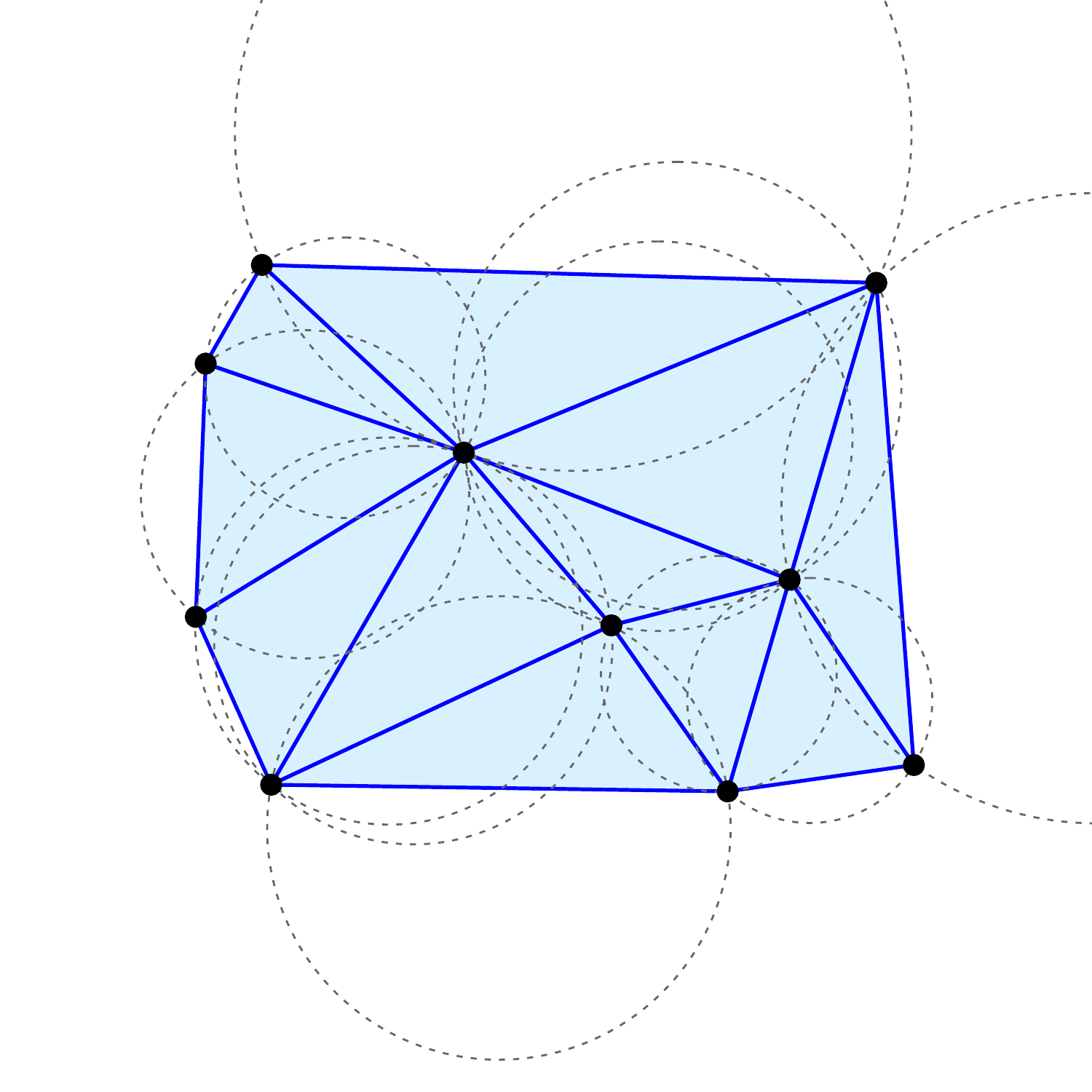}
\caption{Delaunay triangulation with circumcircles (dotted).}
\label{fig:fig_1}
\end{figure}

\section{Background}
\label{sec:Background}
{\subsection{Combinatorial Optimization Problems}}
{The goal of COPs is to identify the optimal object from a finite set of objects to minimize or maximize some objective functions. We consider three types of COPs involving geometry: DT, convex hull, and TSP.}\\

 {\noindent \bf  {DT.} }
DT finds a triangulation such that no point in $P$ is inside the circumcircle of any triangle when a point set $P$ is given~\cite{1}. Figure~\ref{fig:fig_1} presents one example of DT with circumcircles. DT maximizes the minimum angle of all angles of the triangle in the triangulation~\cite{1}. We considered the DT problem to be a COP because DT aims to find a combination of triangles that satisfies the Delaunay condition, which maximizes the minimum angle of the triangles, given a point set. \\

{\noindent \bf  {Convex Hull.} }
Given a set of $n$ points in the 2D plane, the convex hull problem aims to select some or all of them to form a convex polygon such that the area of the resulting polygon is maximized. \\



{\noindent \bf  {TSP.} }
We consider a planar symmetric TSP that is defined as follows: “Given a list of nodes (cities), find the shortest possible route (tour) that visits each node once and returns to the original node”~\cite{2}. Given a planar point set $P=\left \{P_1,\dots,P_n  \right \}$, our goal is to find a permutation of nodes (tour), $\pi$, that visits each node once and has the minimum total length, which is defined as:
\begin{equation}
\label{eqn:eq_9}
L\left (\pi,P  \right ) = {\left \| P_{\pi_n} - P_{\pi_1}  \right \|}_2 + \sum_{i=1}^{n-1}{{\left \| P_{\pi_i} - P_{\pi_{i+1}}  \right \|}_2},
\end{equation}
where ${\left \| \cdot  \right \|}_2$ denotes the L2-norm.

\subsection{Seq2seq Models}
The RNN-based seq2seq models are efficient neural networks for the task of corresponding one sequence to another and are widely used in machine translation~\cite{18,19}. The typical seq2seq models have two RNNs called the encoder and decoder. The encoder reads the input sequence sequentially to create a context vector, and the decoder receives the context vector and sequentially outputs the output sequence.

For the input sequence $P$ and the corresponding output sequence $O^{P}=\left ( O_{1}, O_{2},...,O_{n} \right )$, the seq2seq model estimates the probability of sequence $O^{P}$ using the chain rule as follows:
\begin{equation}
\label{eqn:eq_1}
p\left ( O^{P}|P;\theta  \right )=\prod_{i=1}^{n}p(O_i|O_1,...,O_{i-1},P;\theta),
\end{equation}
where $\theta$ is a learnable parameter. During the inference process, in decoder step $i$, the model predicts the sequence by selecting $O_i$, which maximizes $p(O_i|O_1,...,O_{i-1},P;\theta)$. 

 However, this seq2seq model contains all information in a context vector with a fixed size, resulting in the loss of information, and an attention mechanism is used to solve this problem~\cite{20}. The basic idea of the attention mechanism is that the decoder predicts the output and attention for the entire input sequence of the encoder once  again at every time step. However, even with the attention mechanism, there is still a problem that the output dictionary depends on the length of the input sequence~\cite{21}.

\subsection{Pointer Network (Ptr-Net)}
In the existing seq2seq model, the model is difficult to use for problems where the size of the output dictionary changes according to the input size because the conditional probability, $p(O_i|O_1,...,O_{i-1},P;\theta)$ is calculated and selected from a fixed dictionary. Similar to the seq2seq model, the Ptr-Net consists of two RNNs: an encoder and a decoder. For each time step, the encoder takes an element of the input sequence as input and outputs the embedding at that time. The decoder outputs a pointer to the input data from the embedding received from the encoder through a modified attention mechanism. The Ptr-Net calculates the probability as follows using the modified attention mechanism:
\begin{equation}
\label{eqn:eq_2}
\begin{aligned}
u_j^i  &=v^{T} \mathrm{tanh}  \left (W_1e_j+W_2h_i   \right ), j\in \left ( 1,\cdots,m \right ) \\
p\left ( O_{i} | O_{1},...,O_{i-1}, P  \right ) &=\mathrm{softmax} \left (u^{i}  \right ),
\end{aligned}
\end{equation}
where the hidden states of the encoder and decoder are $\left ( e_1,...,e_m \right )$ and $\left ( h_1,...,h_n \right )$, respectively. In addition, $u^i$ is a vector of length $m$ representing each score of the encoder step, and $W_1$, $W_2$, and $v$ are learnable parameters. Unlike previous attention mechanisms that create a new context vector due to softmax $u^i$, the Ptr-Net considers this to be a probability for each input element. In a dictionary of input elements, $u^i$ is a vector of ``pointers'' to the input elements.

\begin{figure*}[t]
\centering
\hbox
{
\subfigure[] {
\includegraphics[width=3cm]{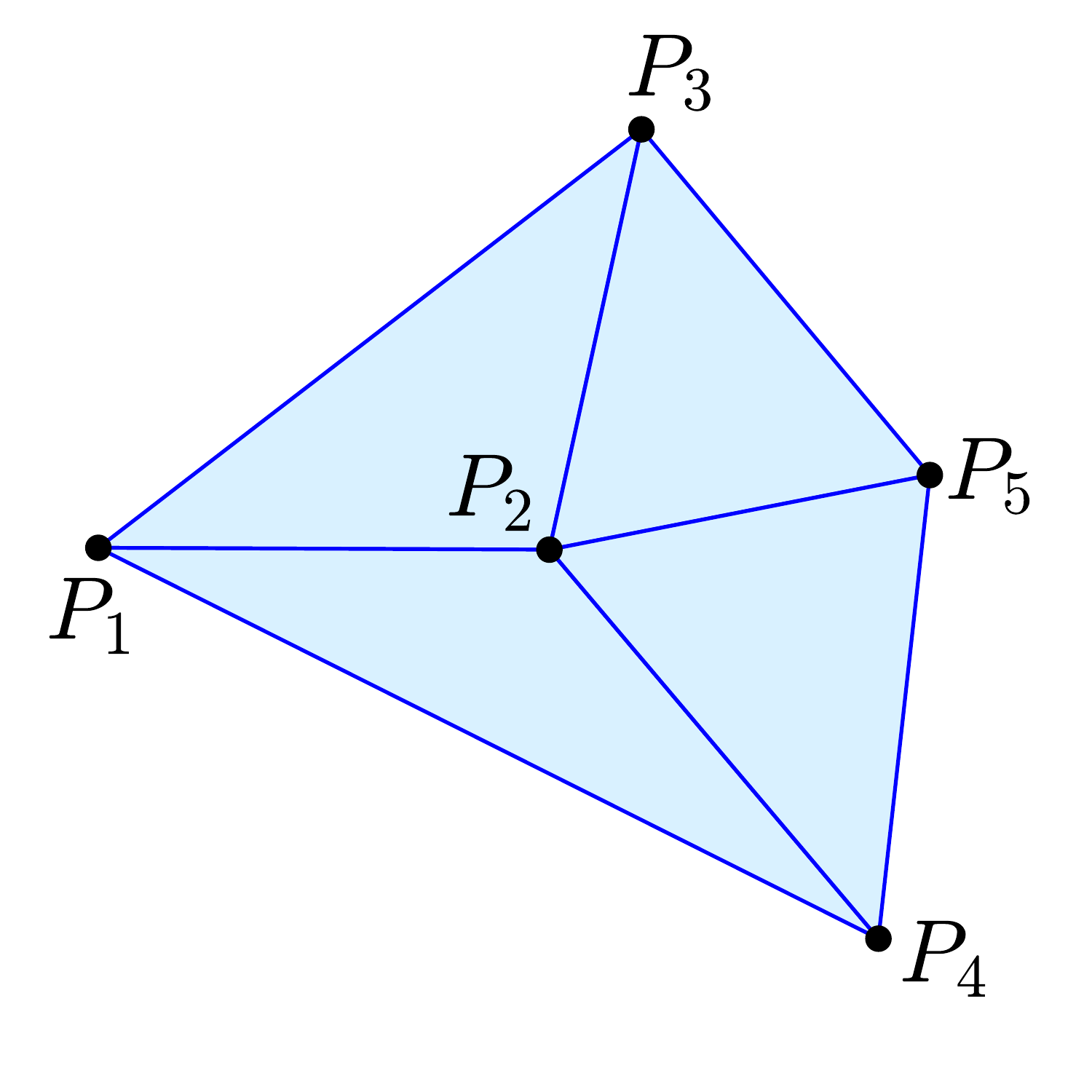}
}
\hspace{0.5cm}
\subfigure[] 
{
\includegraphics[width=8cm]{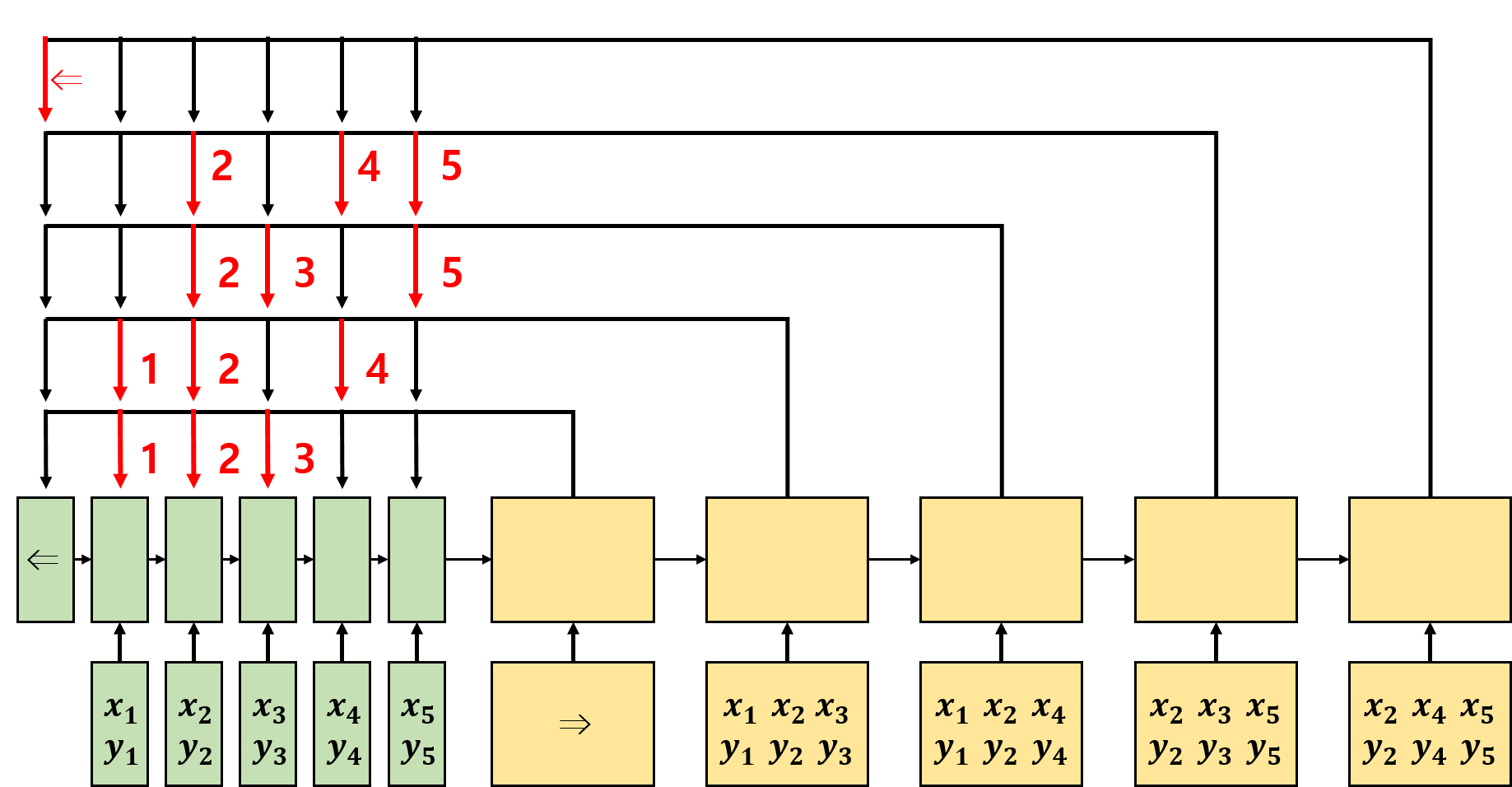}
}
}
\caption{Example using M-Ptr-Net for the DT problem. Input $P=\left \{ P_1,..., P_5 \right \}$ and the output=$\left ( \Rightarrow , \left \{ 1, 2, 3 \right \},\left \{ 1, 2, 4 \right \}, \left \{ 2, 3, 5 \right \},  \left \{ 2,4,5 \right \},\Leftarrow          \right )$. The tokens $\Rightarrow$ and $\Leftarrow$ are beginning and end-of-sequence, respectively. }
\label{fig:fig_3}
\end{figure*}

\subsection{Multi-Ptr-Net}
For the DT problem, three points form a triangle when creating a triangulation. However, the Ptr-Net has limitations in learning the topological relationship of the points, especially when the number of input points is large. Moreover,  the Ptr-Net has a limitation in learning the sequence as a pair; thus, multi-Ptr-Net (M-Ptr-Net) is proposed to improve these limitations~\cite{11}. The M-Ptr-Net is motivated by the multi-label classification method so that multiple embedding results in the encoder can be pointed at simultaneously in one step of the decoder. At each time step of the decoder, it calculates the probability using the following equation:

\begin{equation}
\label{eqn:eq_3}
p\left ( O_{i} | O_{1},...,O_{i-1}, P  \right ) =\mathrm{sigmoid} \left (u^{i}  \right ).
\end{equation}

The original Ptr-Net uses the softmax function for ``pointing'', which is used for the multi-class classification problem. In M-Ptr-Net, the sigmoid function used in the multi-label classification problem is employed instead of the softmax function. Figure~\ref{fig:fig_3} depicts an example of applying M-Ptr-Net to the DT problem. The input includes five planar points, $P_1=\left ( x_1,y_1 \right ),...,P_5=\left ( x_5,y_5 \right )$ with four elements where $\left ( x_i, y_i \right )$ are the Cartesian coordinates of the points. The output sequence represents the solution for the DT problem. Unlike  the Ptr-Net, the strength of the M-Ptr-Net is that it is not affected by the order of the points forming a triangle because it learns the three points forming a triangle as a group.

\subsection{Transformer and Self-attention}
The conventional encoder-decoder models using the attention mechanism have problems in that the amount of learning increases rapidly as the number of steps increases. Moreover, it is challenging to learn the dependence between words in distant locations. To improve these problems, the transformer model is a neural network model that avoids recurrence and relies entirely on the attention mechanism to derive the global dependence of the input and output. The transformer model also uses the encoder-decoder model and includes multi-head attention, self-attention, and positional encoding~\cite{3,23}.

Three types of attention exist in the original transformer model. The first is the attention from the encoder to the encoder, the second is the attention from the decoder to the decoder, and finally, the third is the attention from the decoder to the encoder. Among them, the first two inner attentions are called self-attention, and they occur inside the encoder or decoder. Self-attention is a process of generating a new expression of a corresponding word in consideration of the relative positional relationship of words in a sentence and differs from the recurrence layer and convolutional layer. First, the total computational complexity per layer and the minimum number of sequential operations are both small. Second, it is possible to learn a wide range of dependencies in a short path. Finally, each head presents a model that can be interpreted.

 The transformer interprets the attention as receiving a query and a key-value pair and outputs an output vector that synthesizes the value vector corresponding to the query vector. Single-head attention used in the transformer can be expressed by the following formula~\cite{3,23}:
 \begin{equation}
 \label{eqn:eq_4}
 \mathrm{Head=Attention}\left ( QW^{Q}, KW^{K}, VW^{V} \right )=\mathrm{softmax}\left ( \frac{QW^{Q}\left ( \left ( KW^{K} \right )^{T} \right )}{\sqrt{d_{k}}} \right )VW^{V},
 \end{equation}
 where $Q$ is a  $\left ( q, d_{\mathrm{model}} \right )$ matrix consisting of $q$ query vectors, $K$ is a $\left ( k, d_{\mathrm{model}} \right )$. Moreover, $W^Q$ and $W^K$ are learnable parameters with dimension  $\left ( d_{\mathrm{model}}, d_k \right )$. $W^V$ is a learnable parameter with dimensions $\left ( d_{\mathrm{model}}, d_v \right )$. Self-attention is the case where $Q$, $K$, and $V$ are the same. The result of reinterpreting each embedding is the output considering the relationship between the input embeddings.
 
The authors used the transformer-based model to solve the TSP and enhanced the solution using a simple 2-Opt heuristic~\cite{43}. Kool et al. also proposed the transformer-based model and trained the model using REINFORCE  with a simple baseline~\cite{13}.

\section{Proposed Model}
\label{sec:Model}
The proposed neural network model consists of an encoder and decoder, and each consists of a long short-term memory (LSTM) cell~\cite{24}. The input sequence in the training data is the point set, $P=\left (P_{1},...,P_{m}  \right )$, where $P_i$ is the Cartesian coordinates of $m$ planar points. The outputs, $O^{P}=\left ( O_1,...,O_n \right )$  are sequences representing the solution to the corresponding problem. For the DT problem, each $O_i$ refers to the index of $P$ having an integer value between 1 and $m$, and three consecutive $O_i$ are combined to represent a triangle.
\subsection{Sequence Ordering}
For the seq2seq-based neural network models, the data order has a significant influence on learning performance in the input and output sequences~\cite{2,25}. Sequence ordering is simple and effective in that it only requires changing the sequence order but can significantly improve learning performance. Vinayls et al. found that restricting the equivalence class as much as possible for the output is always better~\cite{2}. The authors proposed a method to perform sequence ordering only for the output sequence. In~\cite{25}, the authors insisted that both the input and output sequence orders are important, but they do not suggest a specific sequence ordering method for the DT problem. The proposed sequence ordering is motivated by the work in~\cite{2,25}.

We propose an efficient input/output sequence ordering method to improve the learning performance of the model. If the sequence is ordered properly and meaningful rules are set, the rules are created in the data, making the model easier to learn. We sorted the input sequence in lexicographic order when putting the input sequence into the encoder. The lexicographic order in the input sequence refers to a method of sorting $P$ in the order of small $x$-axis coordinates with respect to $P$ and sorting in the order of small $y$-axis coordinates if the $x$-axis coordinates are the same. Given that we consider COPs involving geometry, we expect that the model learns much faster and better by lexicographically sorting the points in the input sequence.

For the output sequence, the sequence is ordered as follows. For the DT problem, the two orders in the output sequence are the ordering between triangles and the ordering for the indices of the points that form each triangle. Second, we sort the index of the points forming a triangle by increasing the triangle representation. For example, we use $\left ( 1, 2, 3 \right )$ instead of $\left ( 2, 3, 1 \right )$ to represent a triangle. Finally, we sort each triangle by its incenter coordinates. We compute the incenter coordinates of each triangle and sort them in the order of the small $x$-axis coordinates. If the incenter coordinates of the $x$-axis are the same, we sort them in the order of the small $y$-axis. 

For the convex hull problem, the output sequence starts from the point with the smallest $x$-axis coordinate and is sorted counter-clockwise (CCW). This is because the points corresponding to the convex hull are visited either in the CCW or clockwise (CW) direction. It is known that the optimal planar TSP tour does not have an intersection. Therefore, the output sequence of the TSP is sorted in the same way as the convex hull. For the convex hull problem and the TSP, we observed that there was no noticeable difference in the learning performance of the model irrespective of whether the orientation is CCW or CW.


\begin{figure}[t]
\centering
\includegraphics[width=7cm]{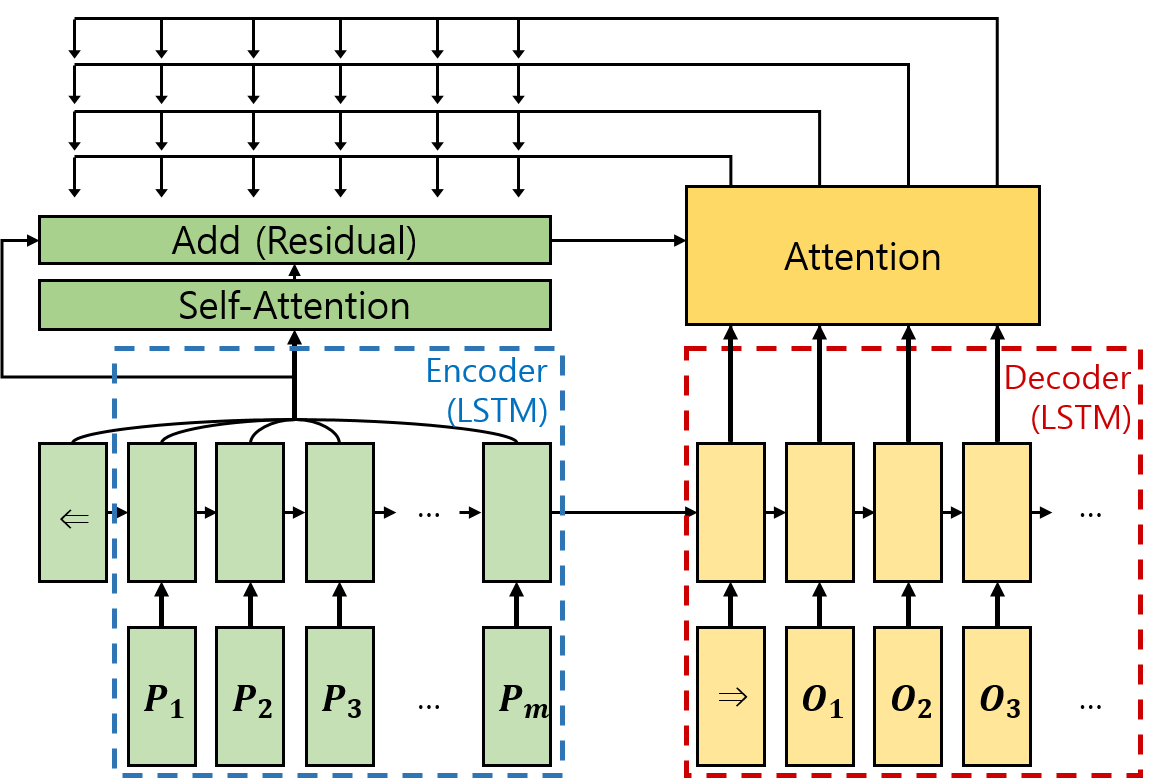}
\caption{Proposed neural net model.}
\label{fig:fig_4}
\end{figure}

\subsection{Attention Mechanism}
\subsubsection{Self-attention}
Self-attention outputs a new embedding reinterpreted in consideration of the relationship between input embeddings as described earlier. Figure~\ref{fig:fig_4} presents the proposed neural network model based on self-attention. In the proposed model, we apply self-attention to the embeddings for each point output of the encoder. We expected that the model can  learn about the relationship and dependence among points better than in the absence of self-attention. We also expected that the model would output a well-analyzed embedding by considering the relationship between points with far distances in the step in the encoder.

In the decoder, self-attention can be used like the transformer model, but we did not use the self-attention mechanism in the decoder for the following two reasons. First, in the encoder, it is difficult to include the relationship between points with far distances. However, in the decoder, the state of the surrounding step is considered  more important than the step with a far distance. Second, because the decoder has a step length of up to 5 to 6 times that of the encoder, the cost of self-attention with a complexity of $O(n^2)$ becomes higher. Additional performance improvement can be achieved if the decoder also uses self-attention, but by using self-attention only in the encoder, the proposed model improves performance within a reasonable computational cost.

\begin{figure}[t]
\centering
\hbox
{
\subfigure[] {
\includegraphics[width=3cm]{./fig3}
}
\hspace{0.5cm}
\subfigure[] 
{
\includegraphics[width=8cm]{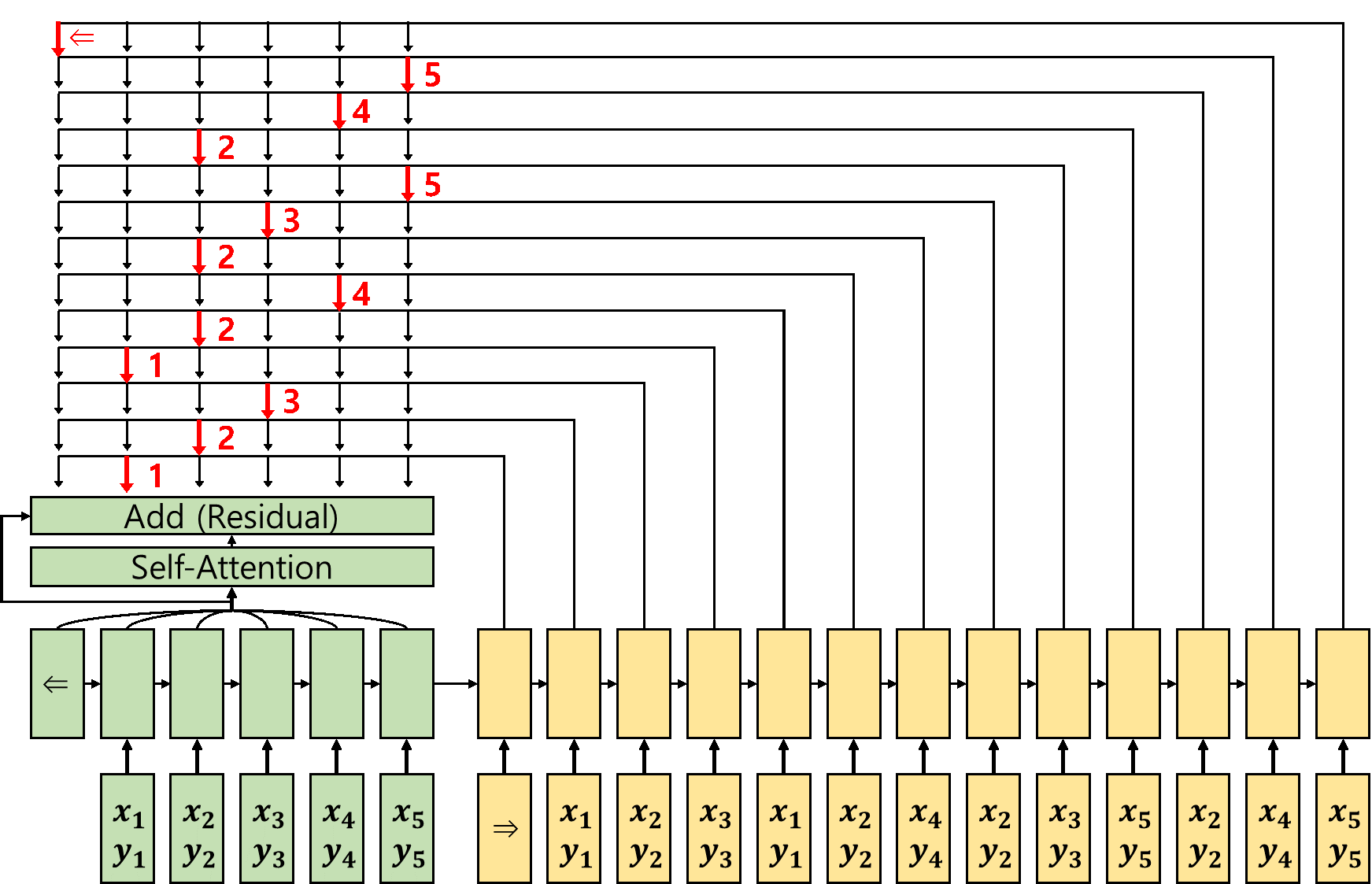}
}
}
\caption{Example of input and output sequence representation for the DT using the proposed model. Input $P=\left ( P_1,...,P_5 \right )$ and $O^{P}=\left ( \Rightarrow, \left ( 1, 2, 3 \right ), \left ( 1, 2, 4 \right ), \left ( 2, 3, 5 \right ), \left ( 2, 4, 5 \right ), \Leftarrow \right )$. The tokens $\Rightarrow$ and $\Leftarrow$ are beginning and end-of-sequence, respectively. }
\label{fig:fig_5}
\end{figure}

Similar to the transformer model~\cite{3,23}, the proposed model also uses the residual connection, but it uses a single head. This is because even a single head creates enough complexity to improve performance greatly.  Figure~\ref{fig:fig_5} displays the model to which the proposed self-attention is applied for DT. Both input/output sequences were sorted using the proposed sequence ordering method. In this example,  $P=\left ( P_1,...,P_5 \right )$ and $O^{P}=\left ( \Rightarrow, \left ( 1, 2, 3 \right ), \left ( 1, 2, 4 \right ), \left ( 2, 3, 5 \right ), \left ( 2, 4, 5 \right ), \Leftarrow \right )$.

\subsubsection{Masking Invalid Outputs}
The conventional attention mechanism does not consider domain knowledge and is thus difficult to apply to geometric COPs. For generating feasible solutions, we use a masking scheme, which sets the attention score functions of infeasible solutions to $-\infty$, such that the model outputs valid solutions. 

The proposed masking scheme for the DT is described in Algorithm 1. First it uses the definition of a triangle: each triangle consists of three points. Therefore, it makes the model generate triangles by forcing the length of the output sequence to be a multiple of three. In addition, at every third step of forming a triangle in the decoder, the function checks whether the Delaunay condition is satisfied for each (candidate) point of $P$. If any candidate point of $P$ violates the Delaunay condition, it masks out that candidate point. The proposed masking scheme is motivated by the definition of DT, and the geometric requirement is determined based on whether it satisfies the Delaunay condition.  

\begin{figure}[t]
\centering
\includegraphics[width=4cm]{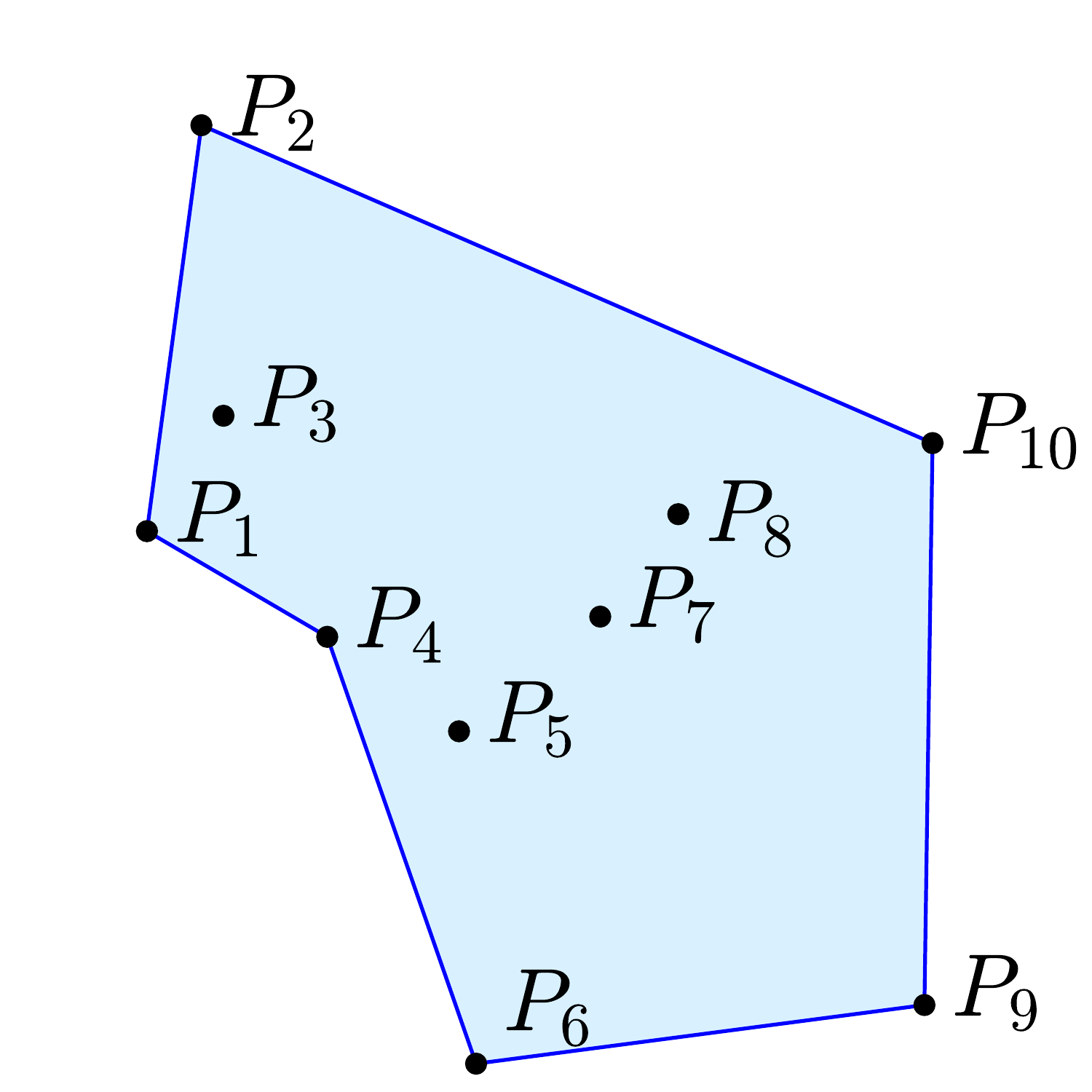}
\caption{Example of the polygon formed by the output sequence, $O^{P}=\left \{ \Rightarrow, 2, 1, 4, 6, 9, 10, \Leftarrow \right \}$, failing to satisfy the convexity requirement. }
\label{fig:fig_new_2}
\end{figure}

For the convex hull problem, we first check whether the candidate points in the output sequence satisfy the convexity requirement. If the orientation (i.e., CW or CCW) of the previous two points and the newly selected candidate point is not the same as the orientation of the first three points in the output sequence, we mask out the candidate points as it does not satisfy the convexity requirement. Figure~\ref{fig:fig_new_2} shows one example where the polygon formed by the output sequence does not satisfy the convexity requirement. Second, the model also masks out the candidate points if they already appeared previously in the output sequence. 

In the TSP, the output tour should visit every city (point) only once and return to the starting city. The model masks out the candidate cities if they appeared previously in the output sequence. Furthermore, if the output sequence does not visit every city once, a high penalty (e.g., $-\infty$) is assigned to the end-of-sequence token such that the output sequence does not produce the  end-of-sequence.

\begin{algorithm}[t]
 \caption{Masking algorithm for the DT}
 \begin{algorithmic}[1]
 \renewcommand{\algorithmicrequire}{\textbf{Input:}}
 \renewcommand{\algorithmicensure}{\textbf{Output:}}
 \REQUIRE Step $i$, input sequence $P$, predicted output sequence $O^{i}=\left ( O_1,...,O_{i-1} \right )$, attention score vector $u^i$, parameter $\gamma$
 \ENSURE  New attention score $\bar{u_j}^i$
    \IF{$i$ mod 3 $\neq$ 1}
        \STATE add $\gamma$ to a score of the end-of-sequence token
    \ENDIF
    \IF{$i$ mod 3 = 0}
        \FOR{$j = 1,..., m$}
        	\STATE triangle $T=\left ( P_{O_{i-2}}, P_{O_{i-1}}, P_{O_{j}}    \right )$
        	\IF{any point of $P$ is inside the circumcircle of $T$}
        		\STATE  $\bar{u_j}^i=-\infty $
        	\ELSE 
        		\STATE $\bar{u_j}^i=u_{j}^{i}$
        	\ENDIF
        \ENDFOR
    \ENDIF
 \end{algorithmic}
 \end{algorithm}

\subsubsection{Beam search decoding}
Beam search is a method for improving efficiency by limiting the number of nodes to be remembered using the best-first search method~\cite{20}. At every time step, the method keeps the most likely sequence elements corresponding to the beam size~\cite{26}. Finally, the decoder selects the sequence with the highest joint probability among the beam subsets. This can be expressed as follows:
\begin{equation}
\label{eqn:eq_6}
 \argmax\limits_{ O^{w}\in O_{B}^{w}} \prod_{i=1}^{w}p\left (O_i | P, O_1,...,O_{i-1}  \right ),
\end{equation}
where $O_{B}^{w}$ is a set of beam candidate sequences at time step $w$.

{\subsection{Training}}

We describe one training step in the training process. A set of input and output (label) sequence pairs, $B=\{(P, O^P )\}$, which is a batch, is given to the model. The model calculates the cross-entropy loss for one training pair $(P, O^P)$ as follows: 

\begin{equation}
\label{eqn:eq_10}
\pazocal{L}\left (P,O^P;\theta  \right )=-\sum_{i=1}^{n}{\log{p\left (O_i|O_1,\dots,O_{i-1},P;\theta  \right ) }}.
\end{equation}
Then, we compute the average over the entire batch to find the loss of the model as follows:
\begin{equation}
\label{eqn:eq_11}
\pazocal{L}\left (\theta  \right )=-\frac{1}{\left |B  \right |}\sum_{k=1}^{\left |B  \right |}{\sum_{i=1}^{n}{\log{p\left (O_i^k|O_1^k,\dots,O_{i-1}^k,P^k;\theta  \right ) }}},
\end{equation}
where $|B|$ is the batch size.

Finally, we update the model parameter $\theta$ to minimize the loss of the model. The model is trained to maximize the probability of the given label sequence conditioned on the input.

\section{Experiments}
\label{sec:Experiments}
{\noindent \bf  {List of Experiments.} }
We conducted experiments to investigate the performance of our model. We ran our tests by increasing the number of points ($m$) in the input point set. The first experiment determines the effect of input/output sequence ordering on model performance. For the proposed neural network model, random ordering and the proposed ordering method described in Section~\ref{sec:Model} are compared. Second, we compared the proposed model with the Ptr-Net~\cite{2}. For this experiment, our goal is to compare the model performance. Therefore, the proposed input/output sequence ordering method is used for both models. For the DT problem, we also compared our model to the M-Ptr-Net~\cite{11} model. The method for multi-label classification was not explicitly mentioned when using M-Ptr-Net. For implementation, the cross-entropy loss was added after the sigmoid function, and then the top three were chosen from the output. There was no mention of input/output sequence ordering for the M-Ptr-Net. Therefore, the same output sequence ordering method mentioned in~\cite{2} was used for the M-Ptr-Net. 

For the TSP, we performed preliminary experiments to determine whether the trainable start token city in the decoder is helpful for performance improvement. The model trained this token city to find the optimal location for the decoder to start. We observed some performance gain for a large-size TSP (i.e., TSP50) when comparing this method with the method of giving a zero vector as input. Therefore, we use this token city strategy for TSP50. 

For the TSP, we compared our model with recent state-of-the-art models: Graph Neural Network (GNN)-based model~\cite{41}, Graph Convolutional Network (GCN)-based model~\cite{14} and Graph Attention Network (GAN)~\cite{13}. In the operational research field, we also compared our model with the A3 algorithm~\cite{2} and popular google OR-Tools~\cite{40}. The A3 algorithm implements the Christofides algorithm with $O\left ( n^3 \right )$~\cite{2} and guarantees that the tour length is not higher than 1.5 times compared to the optimal solution~\cite{2}.

In the experiment, we used two different decoders: greedy and beam search (BS). The greedy decoder selects the vertex with the highest probability at every decoding step~\cite{16}. The BS decoder selects a sequence with the highest joint probability among a subset of beams within a beam width size. Our first two experiments employed greedy decoders, and a BS decoder was used for the third experiment. When using BS, a beam width of four was used in all tasks. The GAN-based model is a reinforcement learning-based model, and it finds the best of four sampled solutions instead of using beam search~\cite{13}. For the TSP, as suggested by authors in~\cite{14}, a beam search strategy that selects the shortest tour among the final beam candidate sequences was also tested. All models were implemented using the Tensorflow 2.0 library, and experiments were conducted using a single Intel Xeon Gold 6152 CPU 2.10 GHz and a single Nvidia Titan V100 PCIe 32 GB GPU.  \\


\begin{figure}[t]
\centering
\hbox
{
\subfigure[DT] {
\includegraphics[width=3.7cm]{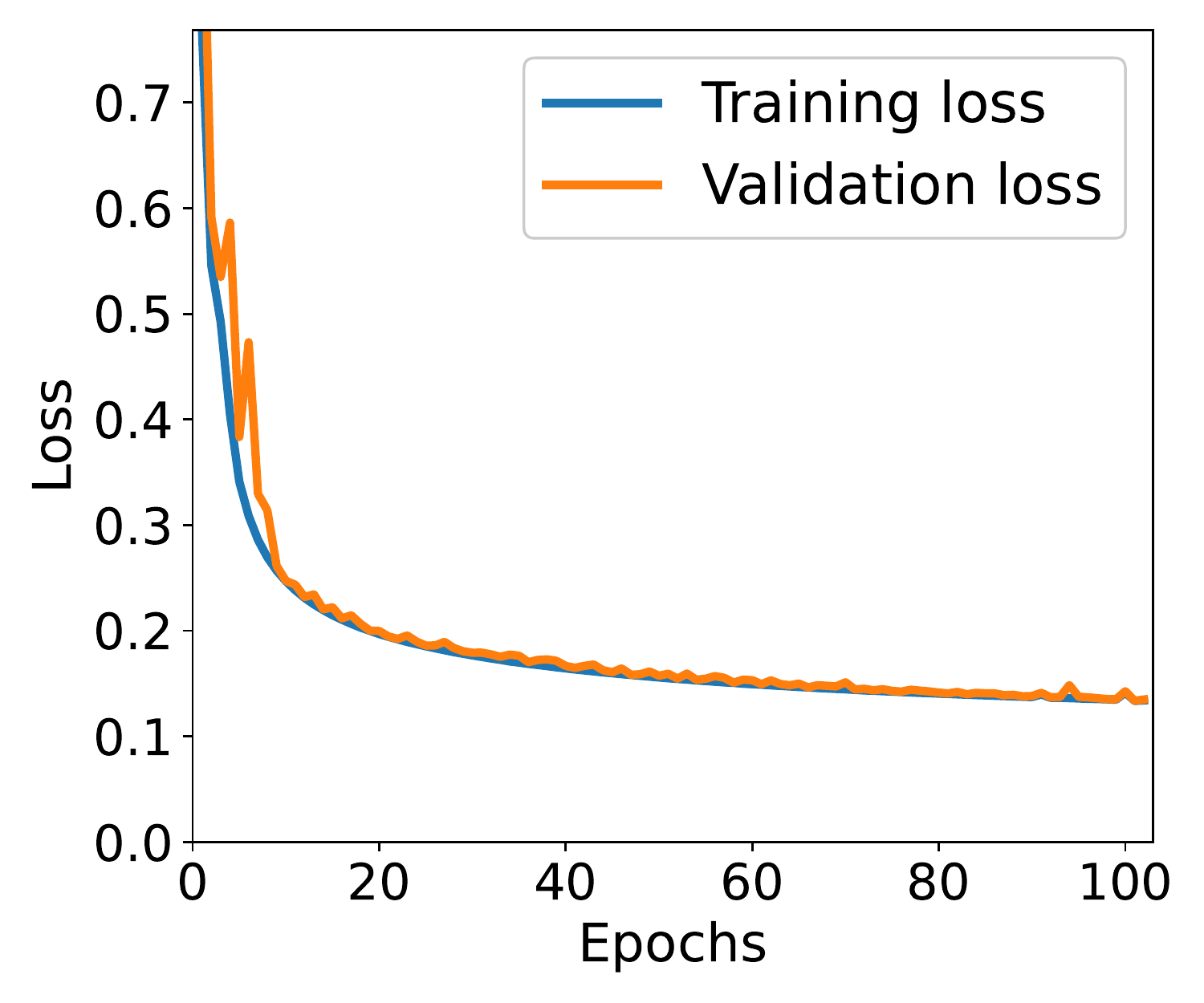}
}
\subfigure[Convex hull] 
{
\includegraphics[width=3.7cm]{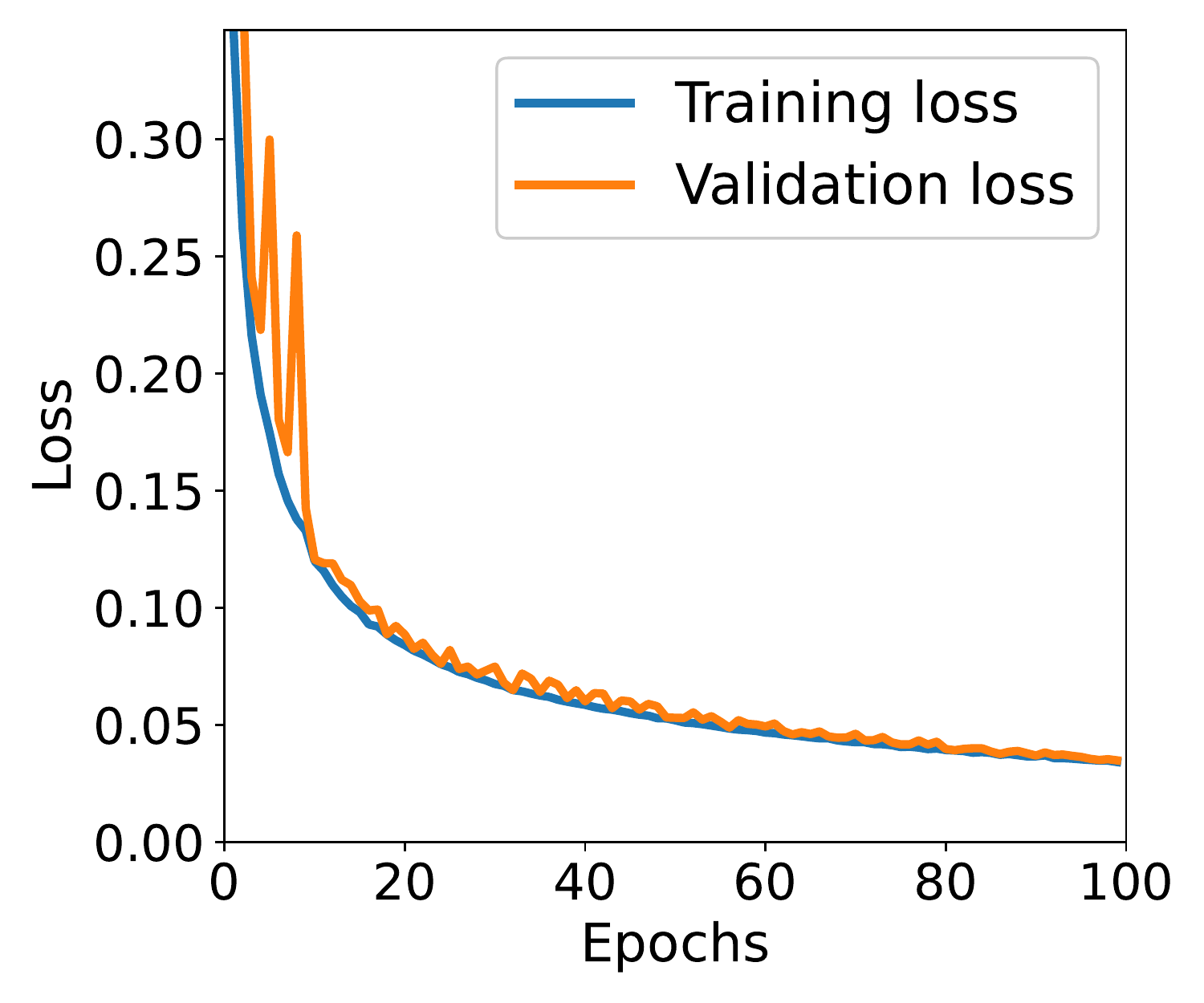}
}
\subfigure[TSP] 
{
\includegraphics[width=3.7cm]{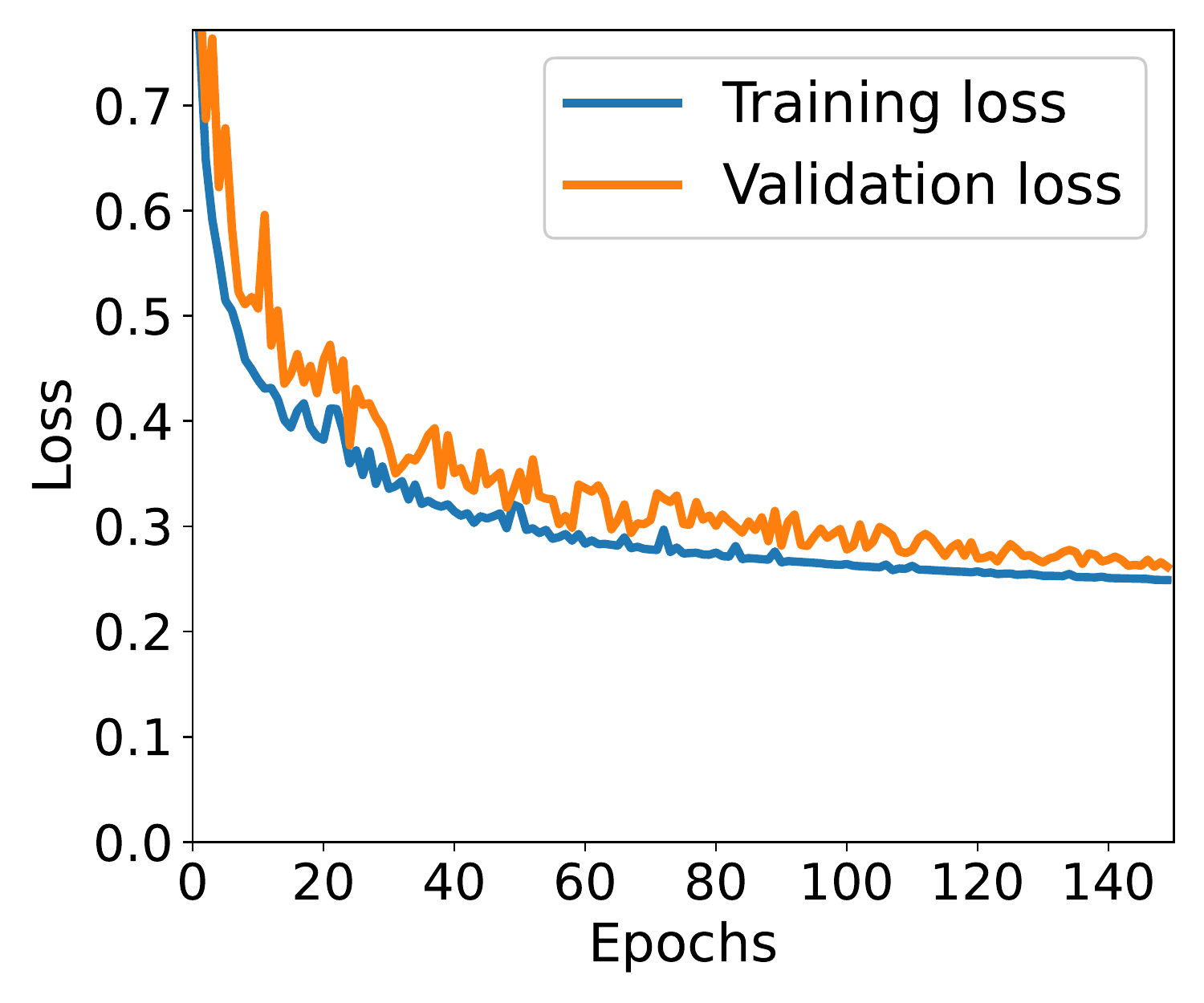}
}
}
\caption{Training loss and validation loss of the proposed model in each problem. (a) DT ($m=20$) (b) Convex hull ($m=100$), (c) TSP ($m=50$). }
\label{fig:fig_loss}
\end{figure}

{\noindent \bf  {Architecture and Hyper-parameters.} }
We used the same architecture and hyper-parameters with the same neural net models throughout all experiments and datasets. Tanh was used as an activation function for the encoder and decoder of the models, and a single-layer LSTM with 256 hidden units was used. The decoder attention mechanism also had 256 hidden units. The Adam optimizer was used~\cite{27} with a learning rate of 0.002, $\beta_1$ of 0.9, and $\beta_2$ of 0.999, and the Xavier method was employed for parameter initialization~\cite{28}. The training was performed until the training loss converges. We did not tune all hyper-parameters to reach the best performance. We used hyper-parameters with good performance on average in all experiments, and we did not fine-tune hyper-parameters to reach the best performance for a specific experiment. We expect better performance when the hyper-parameters are fine-tuned for specific experiments. The training loss and validation loss of the proposed model in each COP are depicted in Figure~\ref{fig:fig_loss}. We observe similar loss convergence patterns for other $m$ values, but the speed at which the loss converges depends on the value of $m$. We trained our model until the loss converges sufficiently.\\

\begin{figure}[t]
\centering
\hbox
{
\hspace{1cm}
\subfigure[Input $P=\left ( P_1,...,P_{10} \right )$ and $O^{P}=\left ( \Rightarrow, P_1, P_6, P_9, P_{10}, P_2, \Leftarrow \right )$.] {
\includegraphics[width=4cm]{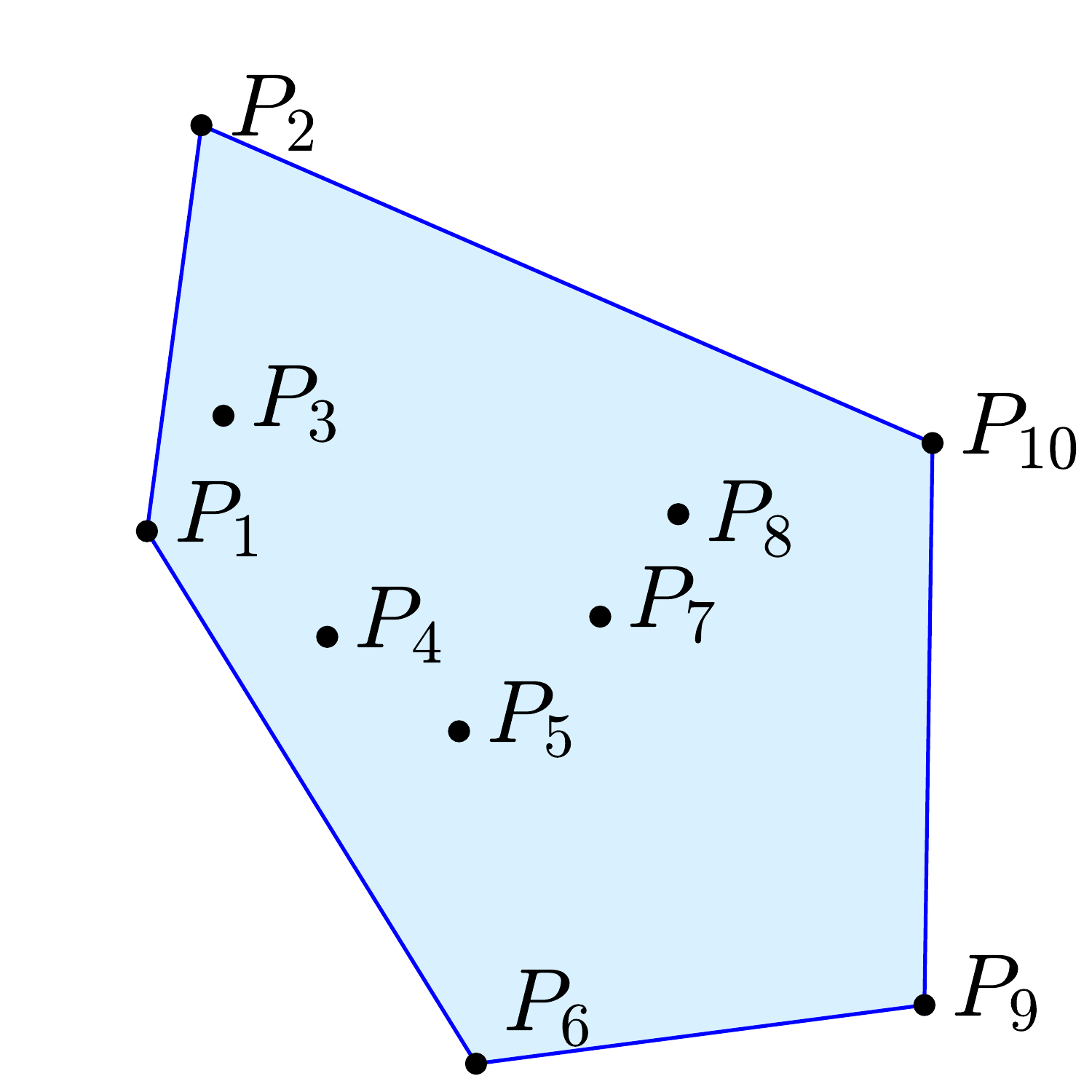}
}
\hspace{1cm}
\subfigure[Input $P=\left ( P_1,...,P_{10} \right )$ and $O^{P}=\left ( \Rightarrow, P_1, P_3, P_2, P_4, P_5, P_7, P_{10}, P_9, P_8, P_6, \Leftarrow \right )$.] 
{
\includegraphics[width=4cm]{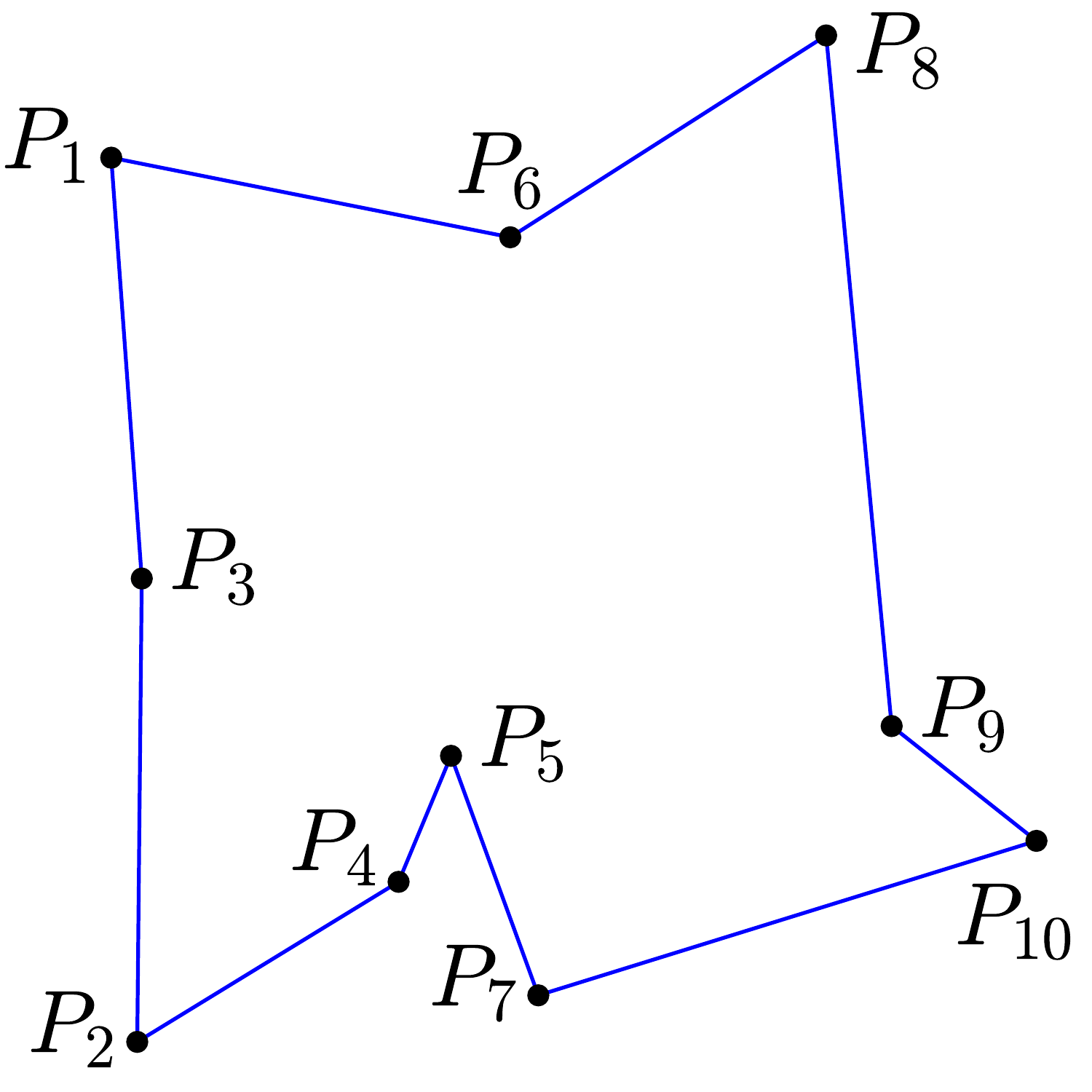}
}
}
\caption{Example of input/output representation for (a) convex hull and (b) TSP. The input sequence is sorted in lexicographic order. The output sequence starts from the point (city) with the smallest x-axis and goes counter clockwise (CCW). The tokens $\Rightarrow$ and $\Leftarrow$ are beginning and end-of-sequence, respectively. }
\label{fig:fig_new_1}
\end{figure}

{\noindent \bf  {Datasets.} }
We generated 1M training example samples (point sets) for each task. Of the total sample data, 90\% of the data were used as the training dataset, and the rest were used for testing. We also tested a k-fold cross-validation approach, but there was no significant difference in performance. For all problems, we sampled from a uniform distribution in $\left [ 0,1 \right ]\times \left [ 0,1 \right ]$. For a given point set, the ground truth data of DT were obtained using MATLAB. For the convex hull and TSP, we used the dataset released in~\cite{2} but reordered the input and output sequences using the proposed ordering method. Figure~\ref{fig:fig_new_1} shows an example of input/output representation for the convex hull and TSP, respectively. For both problems, the proposed input/output sequence ordering method is used. 

For the TSP, we use the Held-Karp algorithm to find the optimal solution for five and ten cities (TSP5 and TSP10)~\cite{2}. When the number of cities is 50 (TSP50), it is extremely costly to produce an optimal solution. Similar to other recent studies~\cite{14,38}, the optimal tour for TSP50 is found using Concorde~\cite{42}. Concorde is considered to produce the fastest optimal TSP solution~\cite{38}. We have released our DT and TSP datasets at https://github.com/hunni10/DelaunayDataset. The convex hull dataset can be found at~\cite{2}.   \\

{\noindent \bf  {Performance Measure.} }
For the DT problem, we evaluated the performance of the proposed neural network model using four different metrics: triangle coverage (TC), accuracy (ACC), triangle count accuracy (TCA) and DT rate (DTR). The first metric is the triangle coverage (TC), which is defined as the ratio of triangles that the model predicts correctly. For two triangular elements consisting of three vertices, even if the permutation of the vertices is different, it is considered as the same triangle. For example, $\left ( 1,2,3 \right )$ and $\left ( 2,3,1 \right )$ represent the same triangle. Additionally, any permutation of the triangles in the output sequence represents the same triangulation. We assumed that the total number of (testing) samples is $S$. We let $\hat{O}_i$ and $O_i$ be the output sequence of the $i^{\mathrm{th}}$ sample of the predictions and the ground truth, respectively. The second metric is accuracy (ACC), which is defined as follows:
\begin{equation}
\label{eqn:eq_7}
\mathrm{ACC}=\frac{\sum_{i=1}^{S}S_{i}}{S},
\end{equation}
where $S_{i}=1$ if $\hat{O}_i$ and $O_i$ represent the same triangulation. Otherwise, $S_{i}=0$. For example, we assume two testing samples each consist of five triangles and that ground truth also has five triangles. For one testing sample, if the model predicts the same triangulation as the ground truth, and for another testing sample, if only four out of five triangles are correctly predicted, then the TC is 0.9, and ACC is 0.5, respectively.

It is also essential to determine whether the number of triangles in the triangulation predicted by the model matches the actual number of triangles in the ground truth. The third metric measures the similarity of the number of triangles predicted by the model to the actual number of triangles in the ground truth. We let $\hat{t}_i$ and $t_i$ be the number of triangles of the predictions and the ground truth of the $i^{\mathrm{th}}$ sample, respectively. The third metric, the triangle count accuracy (TCA), is defined as follows:
\begin{equation}
\label{eqn:eq_8}
\mathrm{TCA}=\frac{\sum_{i=1}^{S}T_{i}}{S},
\end{equation}
where $T_{i}=1$ if $t_i$ and $\hat{t}_i$. Otherwise it is zero. In practice, even the length of the output sequence predicted by the model may not be a multiple of three. If the length of the output sequence $\hat{O}_i$ is not a multiple of three, the output exceeding a multiple of three is excluded when calculating $\hat{t}_i$ because it cannot form triangles during triangulation. For example, if $\hat{O}_i=\left (O_1, O_2, O_3, O_4, O_5 \right )$, then $O_4$ and $O_5$ are excluded.

The final metric is the DT rate (DTR), which measures how well each triangle predicted by the model satisfies the Delaunay condition. It is the ratio of triangles that satisfy the Delaunay condition among the triangles predicted by the model. All four metrics for the DT problem have values between 0\% and 100\%, and the values closer to 100\% indicate better results. In terms of solution quality, both TC and ACC are the most critical metrics for evaluating the performance of the model and should be considered together. This is because the accuracy metric is extremely strict (especially for large $m$) in that the accuracy is zero for one sample if even one triangle among the triangles  predicted by the model is different from the ground truth.

For the convex hull problem, we use two metrics: accuracy (ACC) and area coverage (AC). Accuracy is basically the same as in~\eqref{eqn:eq_7}; the only difference is that if $O_i$ and $O_i$ represent the same polygon, $S_i=1$; otherwise, $S_i=0$. We consider two output sequences to be the same if they represent the same polygon regardless of the orientation (CW or CCW). The second metric is the covered area of the true convex hull. The area coverage is the average ratio (percentage) of the area of the polygon formed by the output sequence to the area of the polygon formed by the ground truth. The area coverage is 100\% when the polygon formed by the output sequence is identical to the polygon of the true convex hull. 

For the TSP, we use two metrics: average tour length (ATL) and valid tour rate (VTR). Average tour length is the average value of the tour lengths of the model. We only count the valid tours when we compute the ATL. As far as the tour is valid, lower tour length indicates better output performance. The second metric is the VTR. The model may output invalid tours. For instance, it could not return to the start city or repeat two cities. The VTR is the average ratio (percentage) of valid tours to the total number of tours. The VTR is 100\% when the output tours of the model are all valid. When evaluating the performance of the model, ATL and VTR should be simultaneously evaluated. If the VTR of the output sequence of the model is low, even if the average tour length of the tours is close to the optimal value, the performance of the model is not satisfactory because there exist too many invalid tours.

\begin{figure*}[t]
\centering   
\vbox
{ 
\hbox
{
\subfigure[m=5] {
\includegraphics[width=5.8cm]{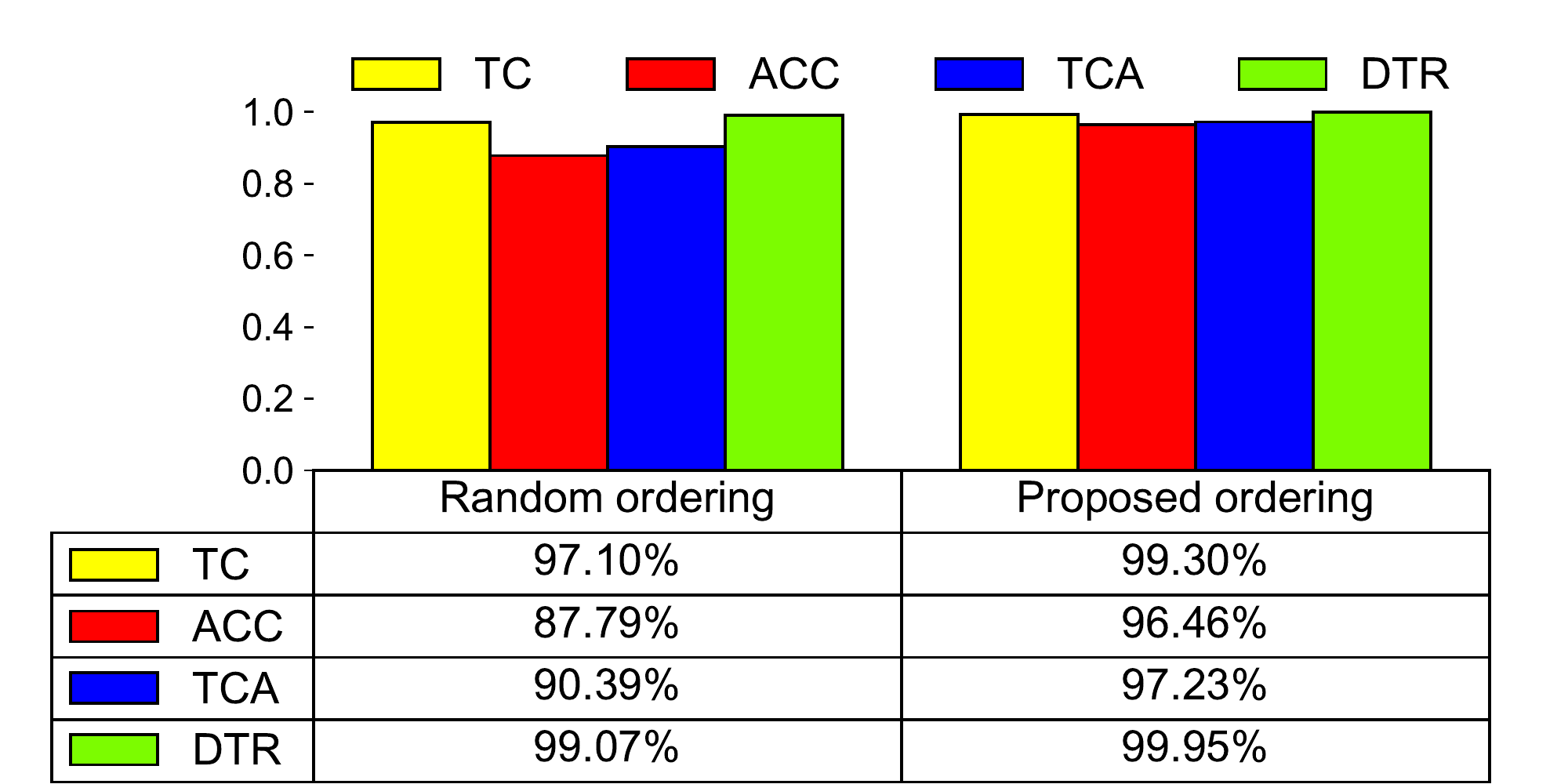}
}
\subfigure[m=10] {
\includegraphics[width=5.8cm]{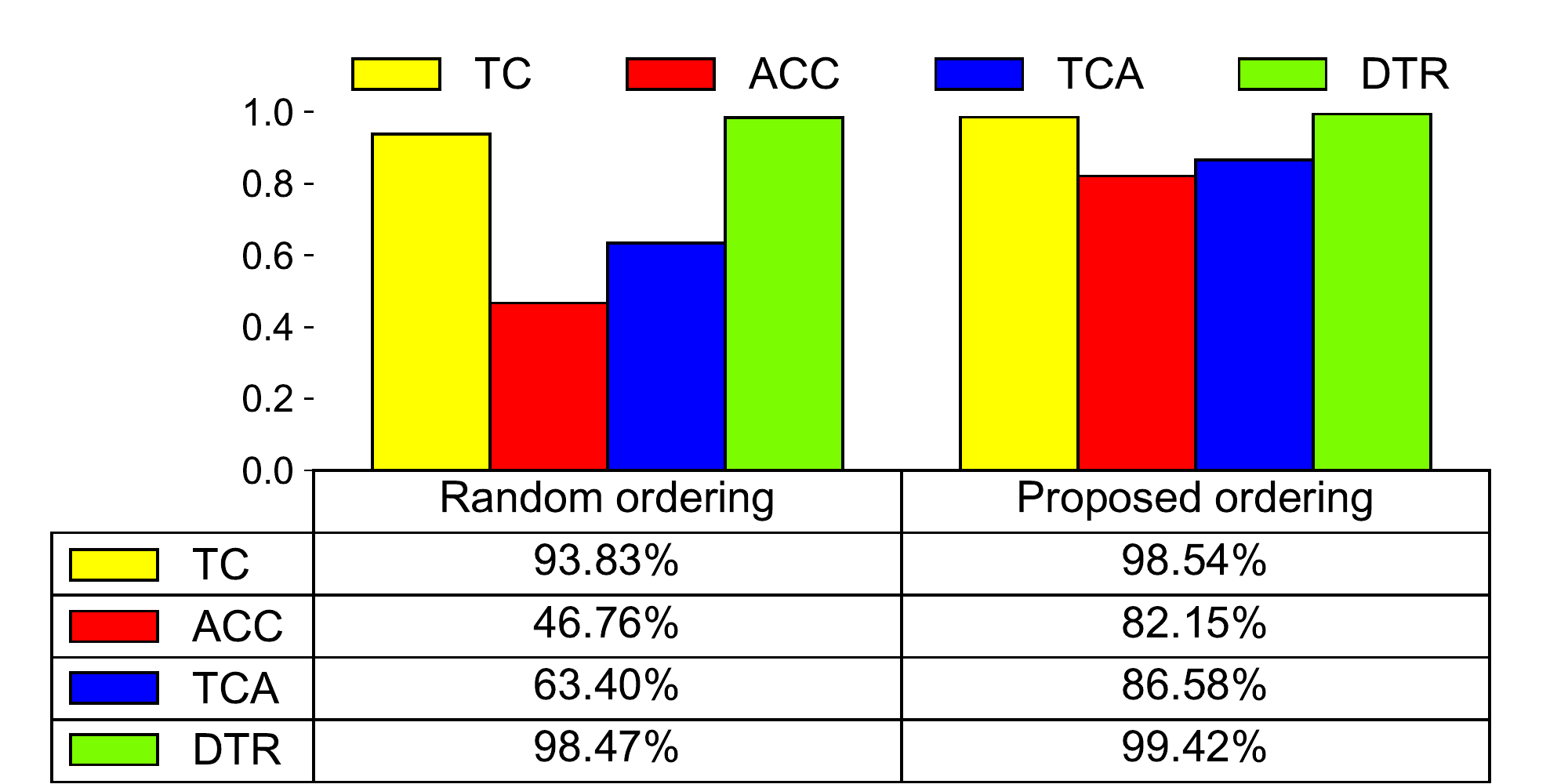}
}
}
\hbox
{
\subfigure[m=15]{
\includegraphics[width=5.8cm]{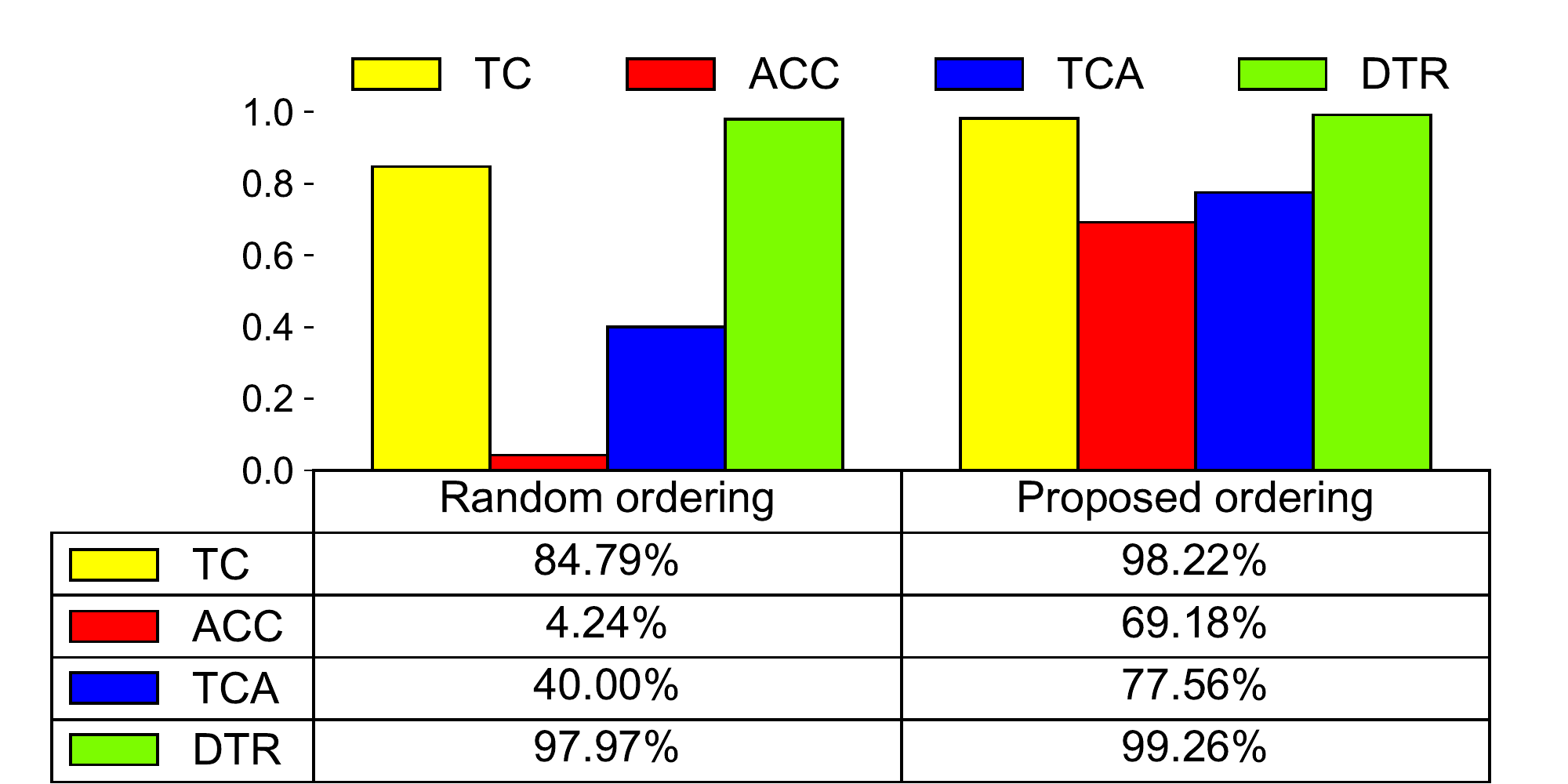}
}
\subfigure[m=20]{
\includegraphics[width=5.8cm]{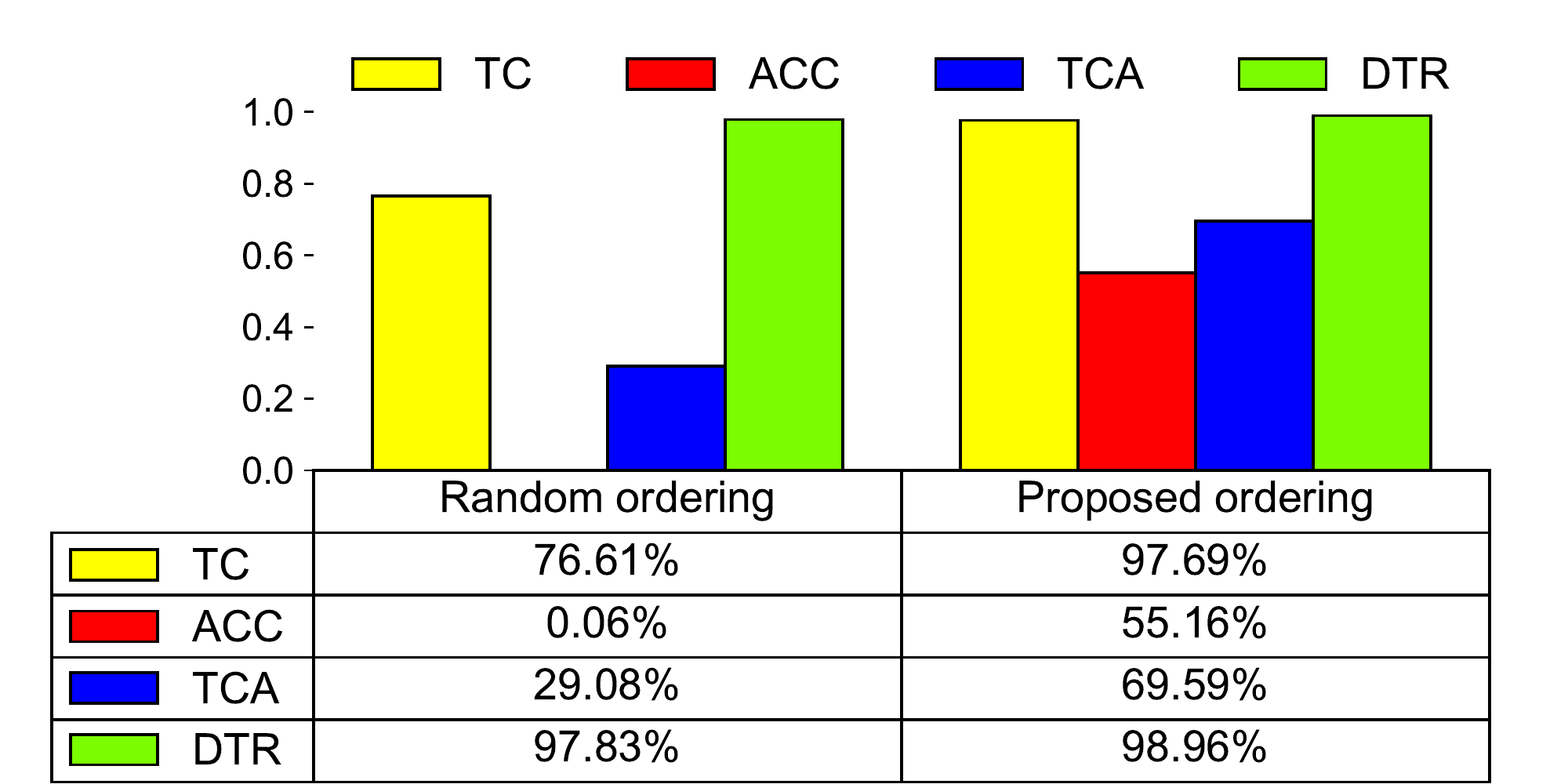}
}
}
}
\caption{Performance of the proposed input/output sequence ordering compared to random ordering for the DT problem.} 
  \label{fig:fig_6}
\end{figure*}

\section{Results}
\label{sec:Results}
\subsection{Experiment 1: Effect of Sequence Ordering}
{\noindent \bf  {DT.} }
Figure~\ref{fig:fig_6} illustrates  the comparison results of the proposed input/output sequence ordering and random ordering. Both methods use the same proposed model, and the only difference is whether the dataset is sorted or not. The performance is much better in all metrics when the sequence was sorted using the proposed ordering method for all $m$ values. This is consistent with the results of previous work in which the sequence order dramatically influences  the performance of the Ptr-Net-based models. Sequence ordering must be performed because performance is greatly improved by simply changing the sequence order of the dataset without making any changes to the model.

As the $m$ value increases, the sorted and random ordering performance gap is much more significant in both TC and ACC metrics. For $m$ = 5, the TC and ACC of the sorted ordering are 99.30\% and 96.46\%, respectively, whereas the TC and ACC of the random ordering are 97.10\% and 87.79\%. When the sequences are not sorted in the case of $m$ = 20, TC is 76.61\%, and ACC is 0.06\%. If the sequences are sorted, TC and ACC are 97.69\% and 55.16\%, respectively.

Interestingly, sorting the input/output sequence also has a significant effect on the TCA. In sorting the input/output sequence for all $m$ values, the number of triangles predicted by the model is more consistent with the actual number of triangles of the ground truth. No significant difference exists between the DTR ratio for all $m$ values, whether the sequences were sorted or not. \\

\begin{figure}[t]
\centering
\hbox
{
\subfigure[m=50] {
\includegraphics[width=6cm]{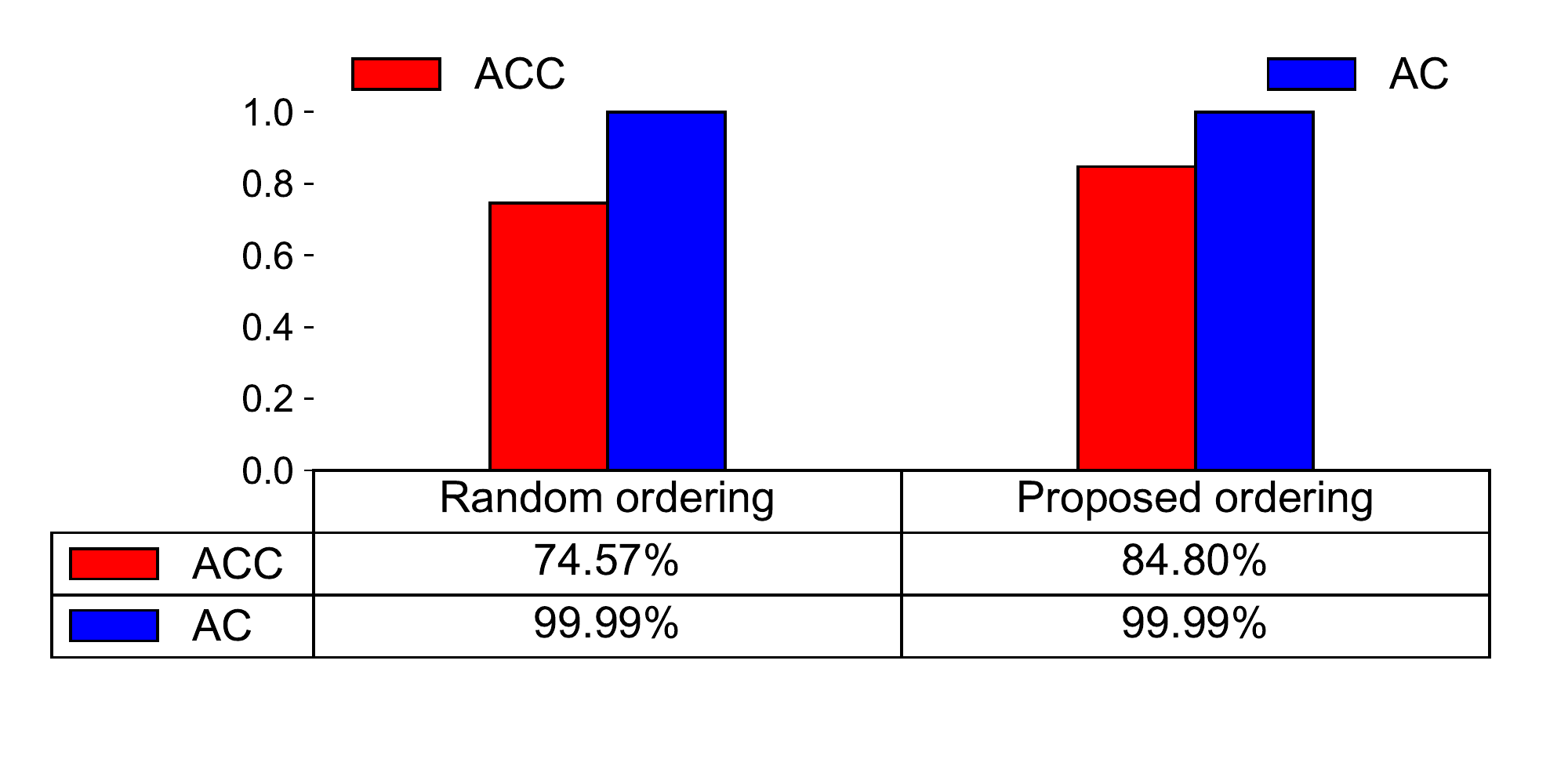}
}
\subfigure[m=100] 
{
\includegraphics[width=6cm]{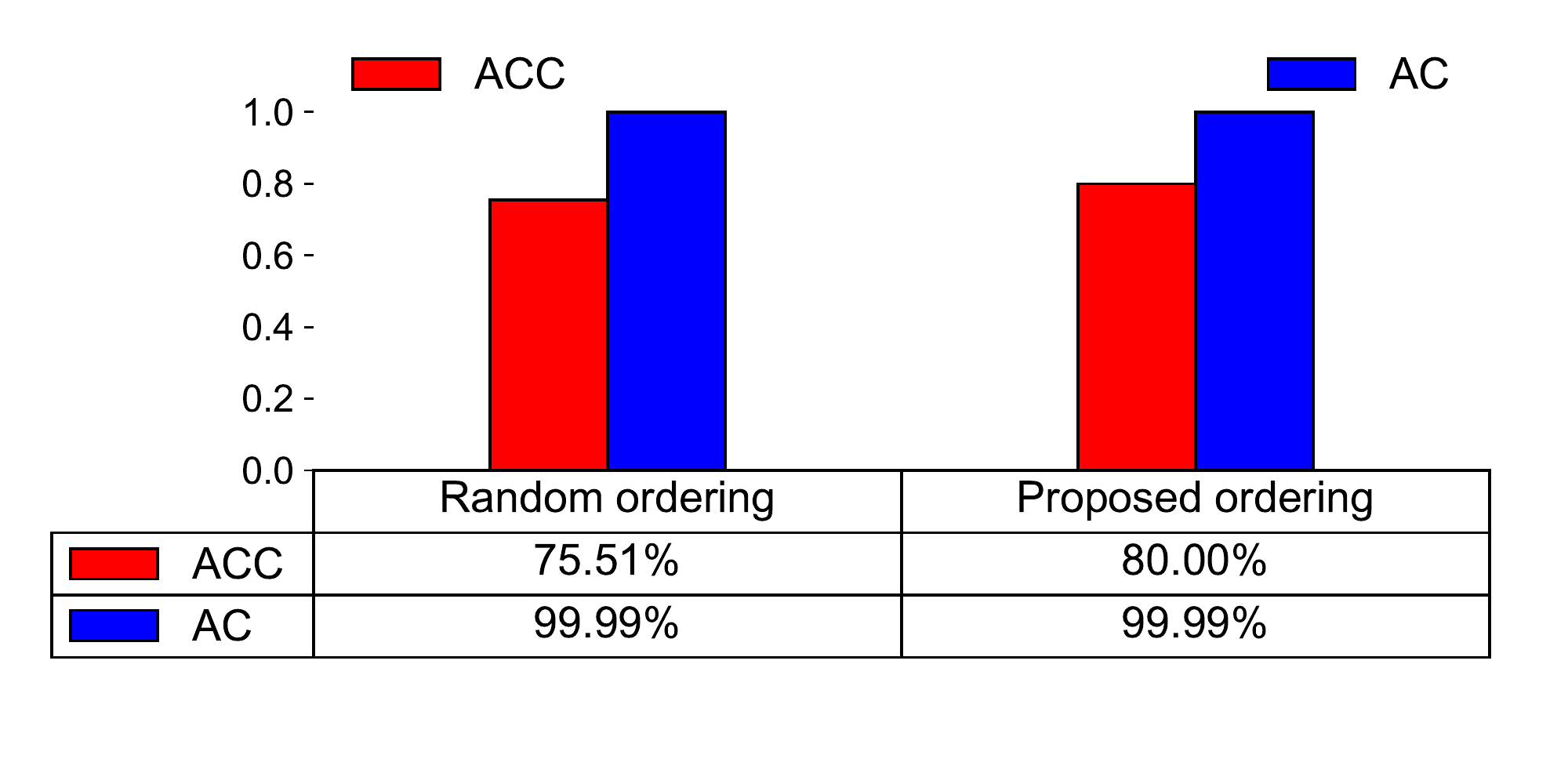}
}
}
\caption{Performance of the proposed input/output sequence ordering compared to the random ordering for the convex hull problem.}
\label{fig:fig_conv_1}
\end{figure}

{\noindent \bf  {Convex Hull.} }
Figure~\ref{fig:fig_conv_1} shows the performance of the proposed input/output sequence ordering compared to the random ordering for the convex hull problem. The proposed ordering is effective and improves the performance for all $m$ values. When $m$ is 50, the proposed ordering shows 84.80\% accuracy and 99.99\% area coverage, while random ordering shows 74.57\% accuracy and 99.99\% area coverage. As $m$ increases, the prediction accuracy of the model decreases, but the area coverage is nearly 100\%, so the predicted polygon is close to the ground truth. \\

\begin{table}[t]
\centering
\begin{tabular} { l || c | c || c | c|| c| c}
\multirow{2}{*}{Methodl}  & \multicolumn{2}{c||}{TSP5} & \multicolumn{2}{c||}{TSP10} & \multicolumn{2}{c}{TSP50} \\       
                                                    &  ATL   & VTR    &  ATL  & VTR    &  ATL  & VTR    \\    
\hline \hline
Optimal & 2.12 & 100\% & 2.87 & 100\%  &  5.69 & 100\% \\
\hline 
Random  & \multirow{2}{*}{2.12} & \multirow{2}{*}{100\%} & \multirow{2}{*}{2.89}& \multirow{2}{*}{100\%}  &  \multirow{2}{*}{6.52} & \multirow{2}{*}{100\%} \\
ordering  &                                    &                                      &                                  &                                        &                                &                                          \\
\hline 
\textbf{Proposed}  & \textbf{\multirow{2}{*}{2.12}} & \multirow{2}{*}{100\%} & \textbf{\multirow{2}{*}{2.88}} & \multirow{2}{*}{100\%}  &  \textbf{\multirow{2}{*}{5.97}} & \multirow{2}{*}{100\%} \\
\textbf{ordering}  &                                    &                                      &                                  &                                        &                                &                                          \\
\hline 
\end{tabular}
\caption{Average tour length (ATL) and valid tour rate (VTR) of the proposed input/output sequence ordering compared to the random ordering. }
\label{tab:tab_1}
\end{table}

{\noindent \bf  {TSP.} }
Table~\ref{tab:tab_1} presents the average tour length (ATL) and valid tour rate (VTR) of the proposed input/output sequence ordering compared to the random ordering.  We observed that the input/output sequence order affects the ATL, while the proposed input/output sequence ordering is effective in improving the ATL. When the number of points (city) is not large, the performance difference between random ordering and proposed ordering was not large, but as the number of cities increases (e.g., TSP50), the proposed ordering showed much better performance than random ordering. That is, the ordering matters more when the sequence length increases. When the proposed model is used, the VTR was 100\% regardless of which ordering method was used.

\begin{figure}[t]
\centering   
\vbox
{ 
\hbox
{
\subfigure[m=5] {
\includegraphics[width=5.8cm]{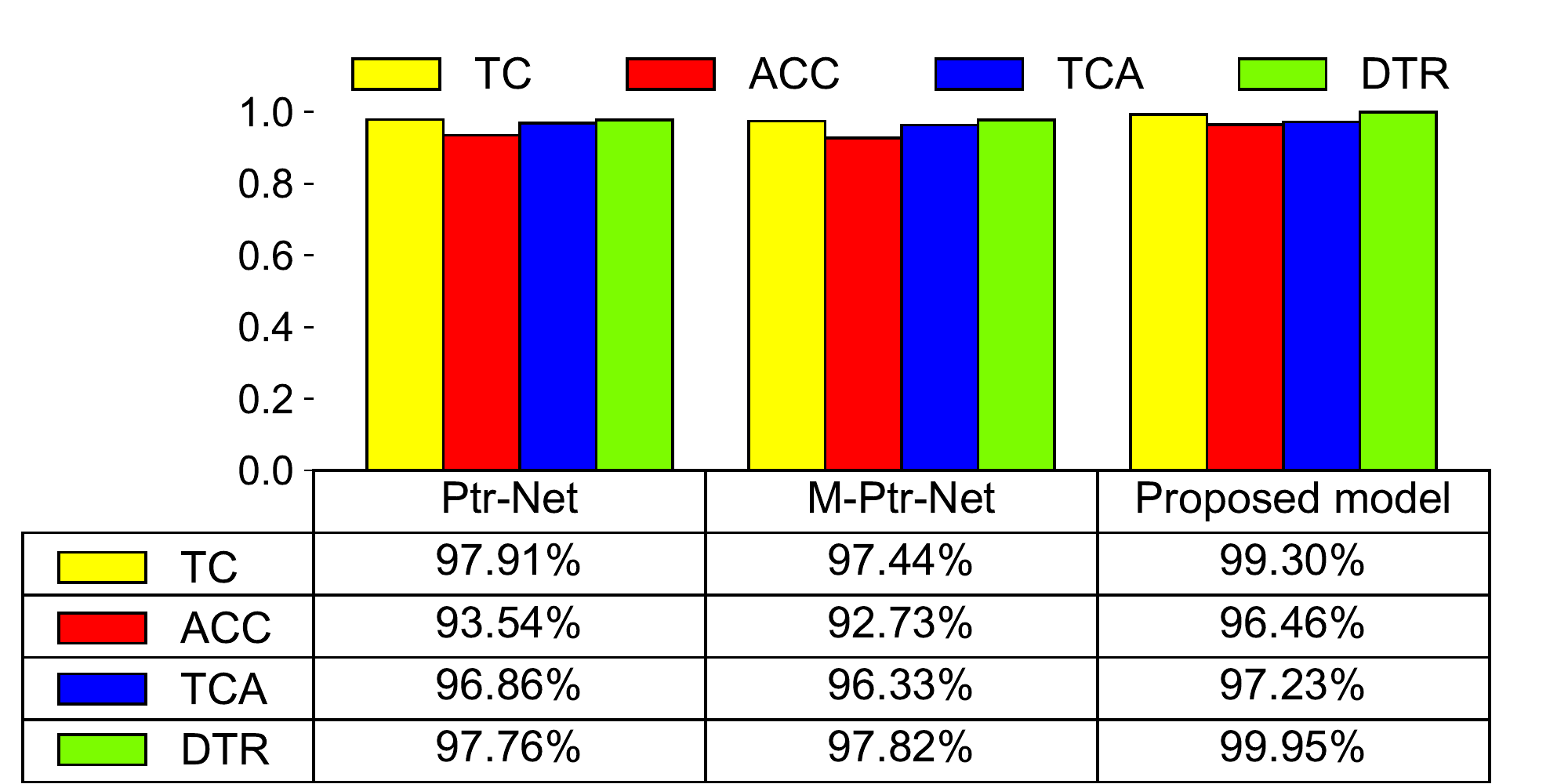}
}
\subfigure[m=10] {
\includegraphics[width=5.8cm]{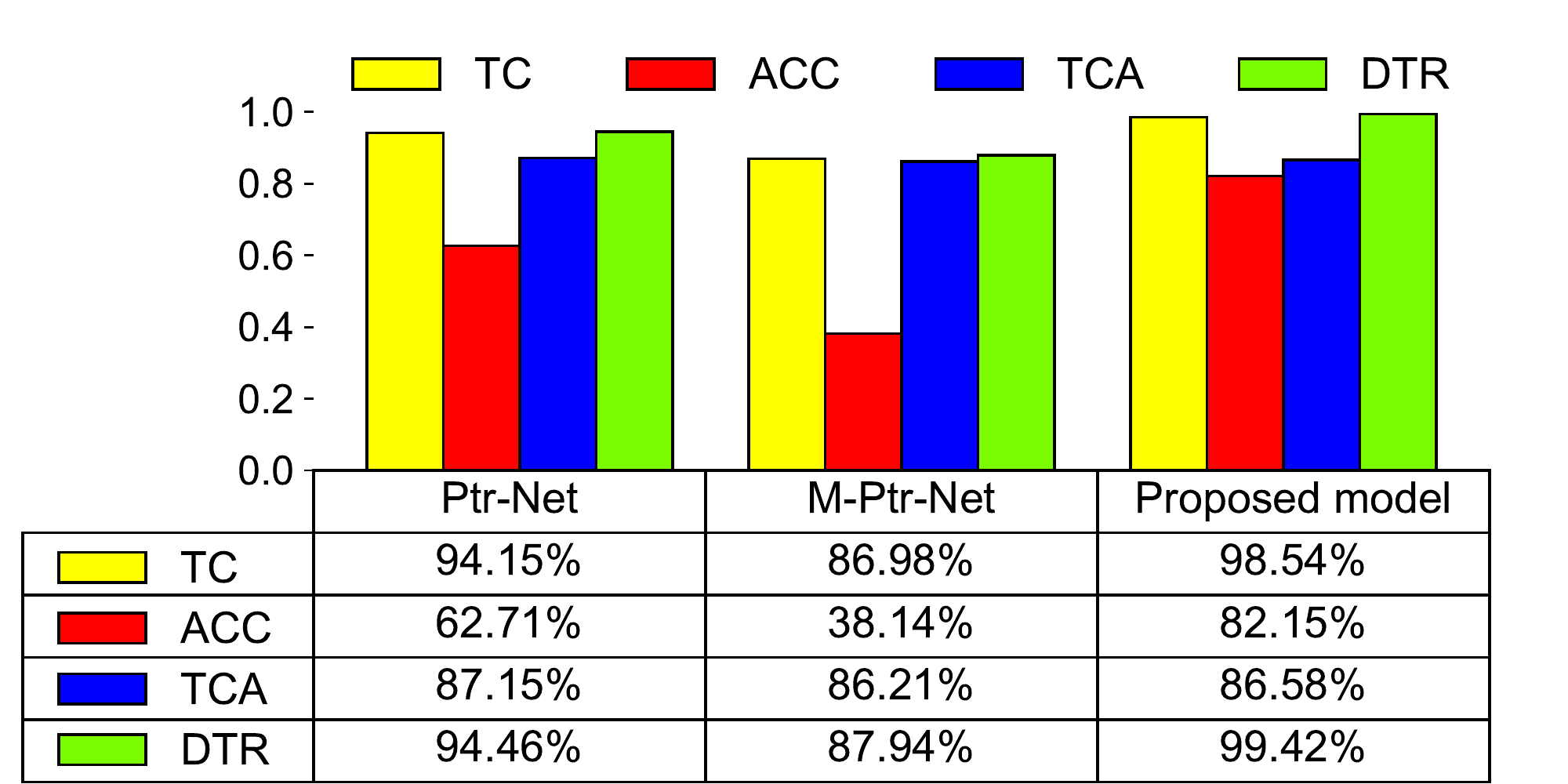}
}
}
\hbox
{
\subfigure[m=15]{
\includegraphics[width=5.8cm]{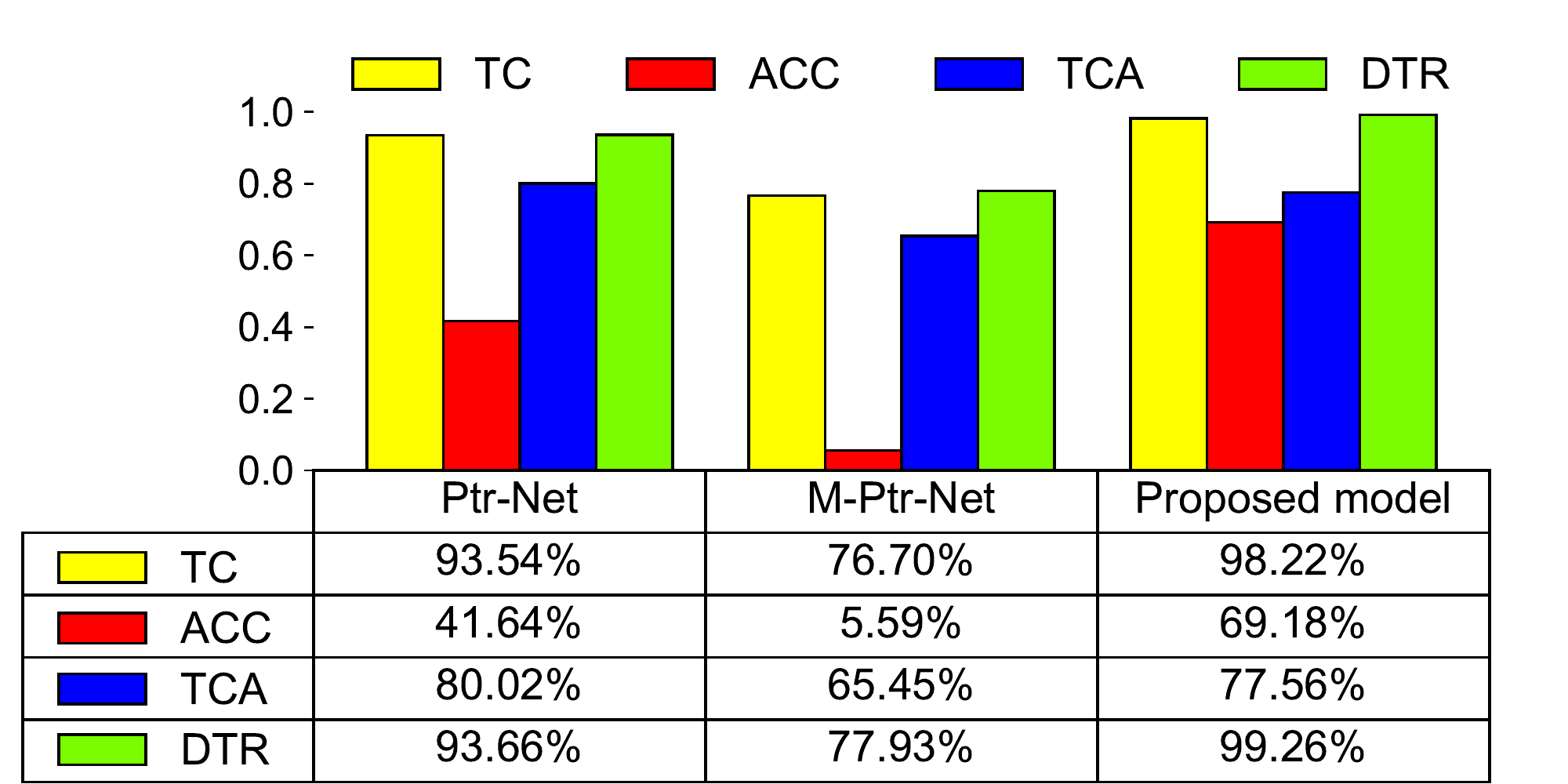}
}
\subfigure[m=20]{
\includegraphics[width=5.8cm]{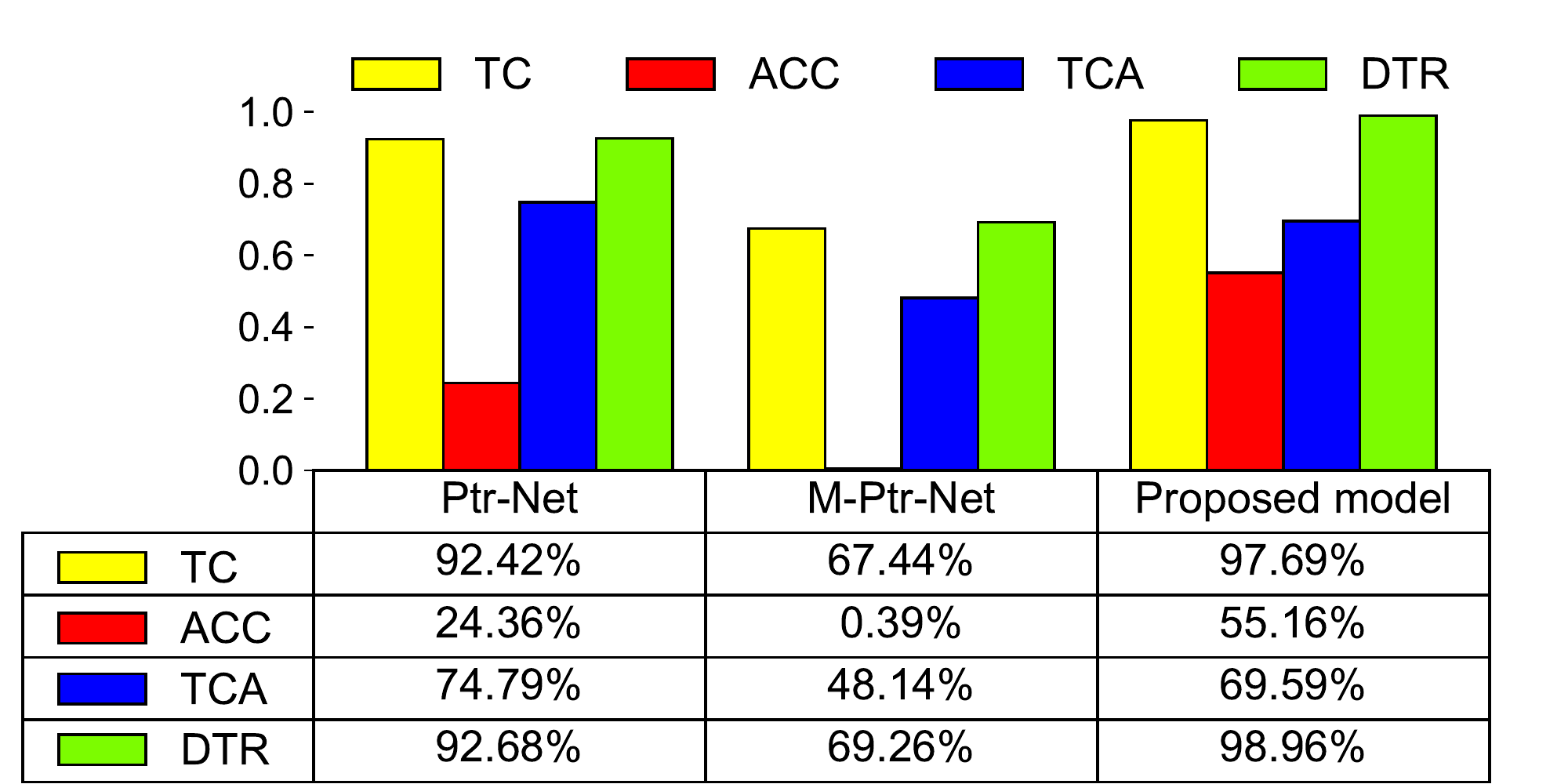}
}
}
}
\caption{Performance of the proposed model compared to the previous work for the DT problem.} 
  \label{fig:fig_7}
\end{figure}

\begin{figure}[t]
\centering
\includegraphics[width=11cm]{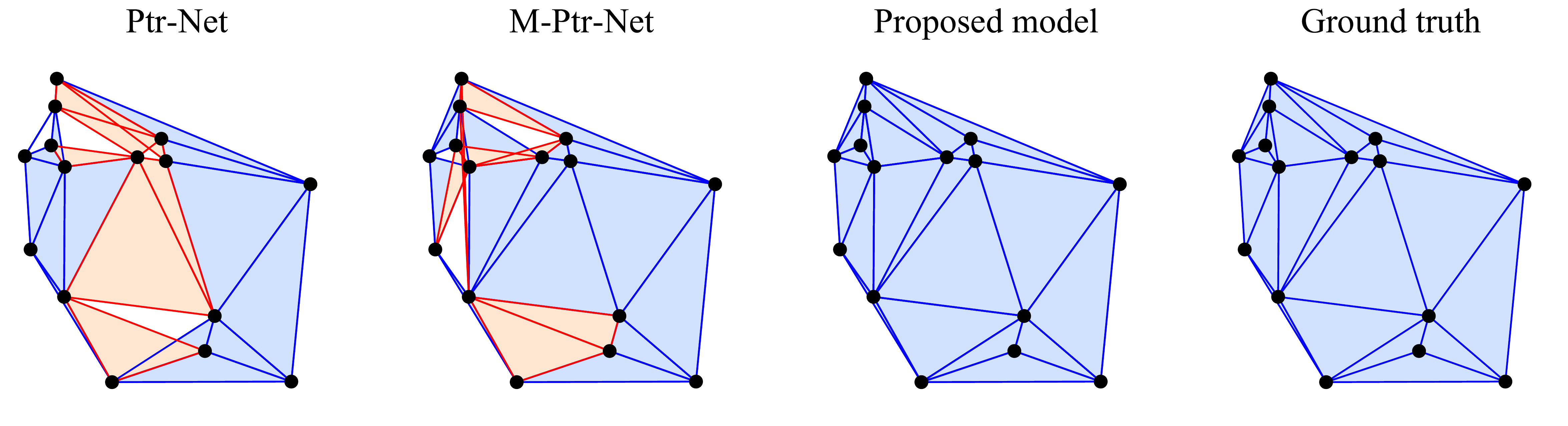}
\caption{Predictions examples for each model with the greedy decoder and the ground truth when $m$ = 15. Blue elements represent elements in the ground truth, and red elements represent elements not in the ground truth. Only the proposed model predicts the ground truth.}
\label{fig:fig_8}
\end{figure}

\begin{figure}[t]
\centering
\hbox
{
\subfigure[m=50] {
\includegraphics[width=6cm]{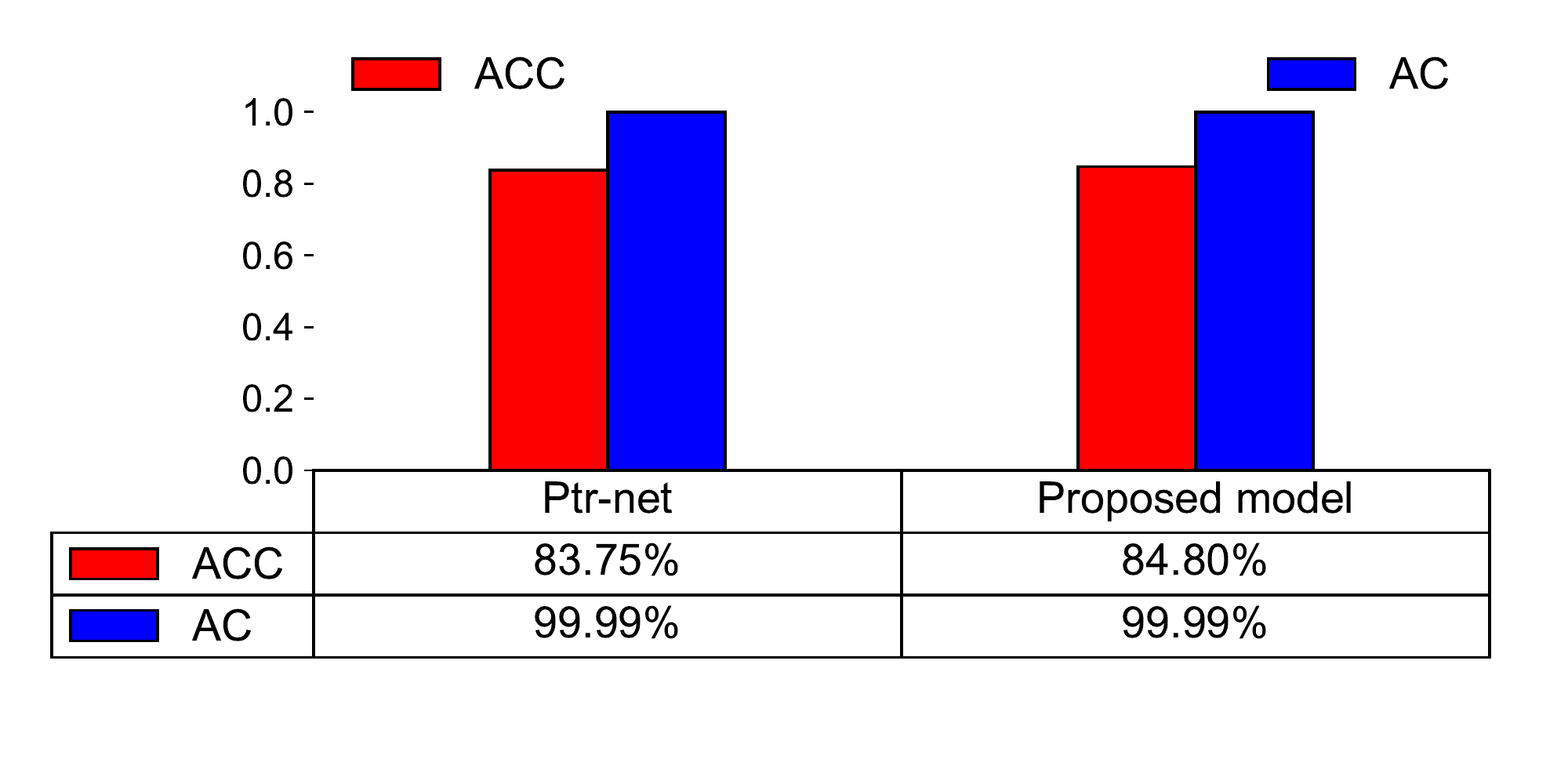}
}
\subfigure[m=100] 
{
\includegraphics[width=6cm]{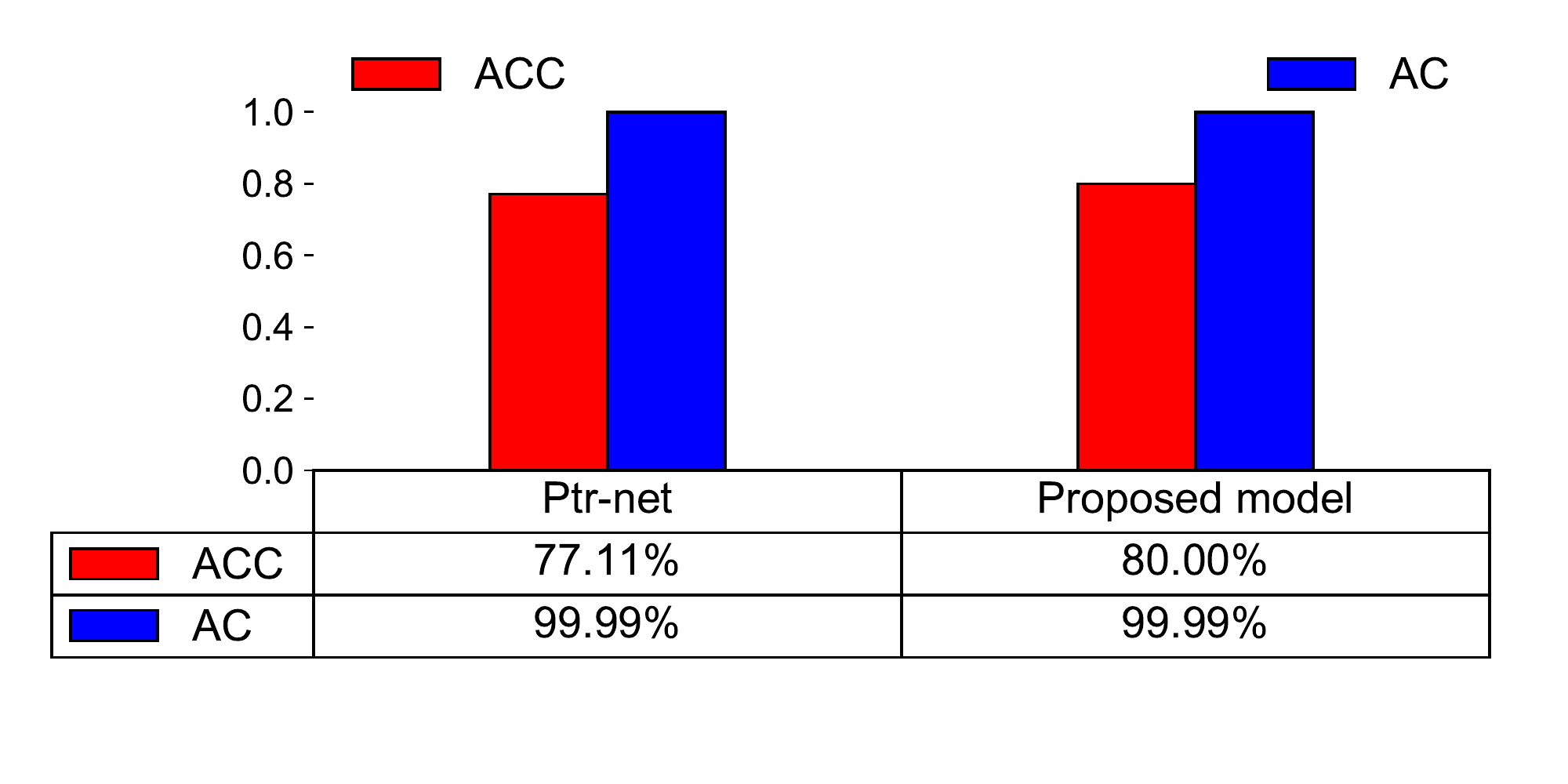}
}
}
\caption{Performance of the proposed model compared to the Ptr-Net for the convex hull problem.}
\label{fig:fig_conv_2}
\end{figure}

\subsection{Experiment 2: Comparison with Previous Work}
{\noindent \bf  {DT.} }
Figure~\ref{fig:fig_7} depicts the results of the proposed model compared with the existing deep learning-based DT generation methods according to $m$ values. We observe that the performance of the proposed model outperforms other methods for all $m$ values in the TC, ACC, and DTR. This performance is primarily  for two reasons. First, the model can learn the point-to-point relationship through self-attention in the encoder. Second, the proposed masking scheme provides a high penalty function to candidates who do not satisfy the Delaunay condition. In particular, the performance gap between the proposed model and other methods increases as the value of $m$ increases. Specifically, the proposed model achieves TC = 98.54\% and ACC = 82.15\% when $m$ is 10, whereas Ptr-Net achieves TC = 94.15\% and ACC = 62.71\%, and M-Ptr-Net achieves TC = 86.98\% and  ACC = 38.14\%. When the value of $m$ is 20, the proposed model achieves TC = 97.69\% and ACC = 55.16\%, whereas Ptr-Net achieves TC = 92.42\% and ACC = 24.36\%, and M-Ptr-Net achieves TC = 67.44\% and ACC = 0.39\%.

We observe that especially for large $m$ values, the length of the output sequence predicted by the Ptr-Net is slightly longer than that of the ground truth in many cases. In these cases, the TCA may appear to be high because it excludes outputs exceeding a multiple of three from the output sequence. Even if the predicted sequence length is not exactly correct, only the length of the valid sequence, which corresponds to the number of valid triangles, is only considered.

Among the compared models, the M-Ptr-Net has the worst performance for all metrics. In particular, the ACC becomes zero when $m$ = 20. Unlike the Ptr-Net, the M-Ptr-Net chooses three points simultaneously, so the model does not sufficiently learn the relationship among the three points to be selected. Figure~\ref{fig:fig_8} displays one example of the predicted triangulation of each model and the ground truth when $m$ = 15. The  blue elements represent elements in the ground truth, and red elements represent elements not in the ground truth. In this example, only the proposed model exhibits the same triangulation as the ground truth, but other models fail to predict the ground truth. We only demonstrate one example, but the proposed model performance is superior to other existing models in most testing samples. \\

\begin{figure}[t]
\centering
\includegraphics[width=10cm]{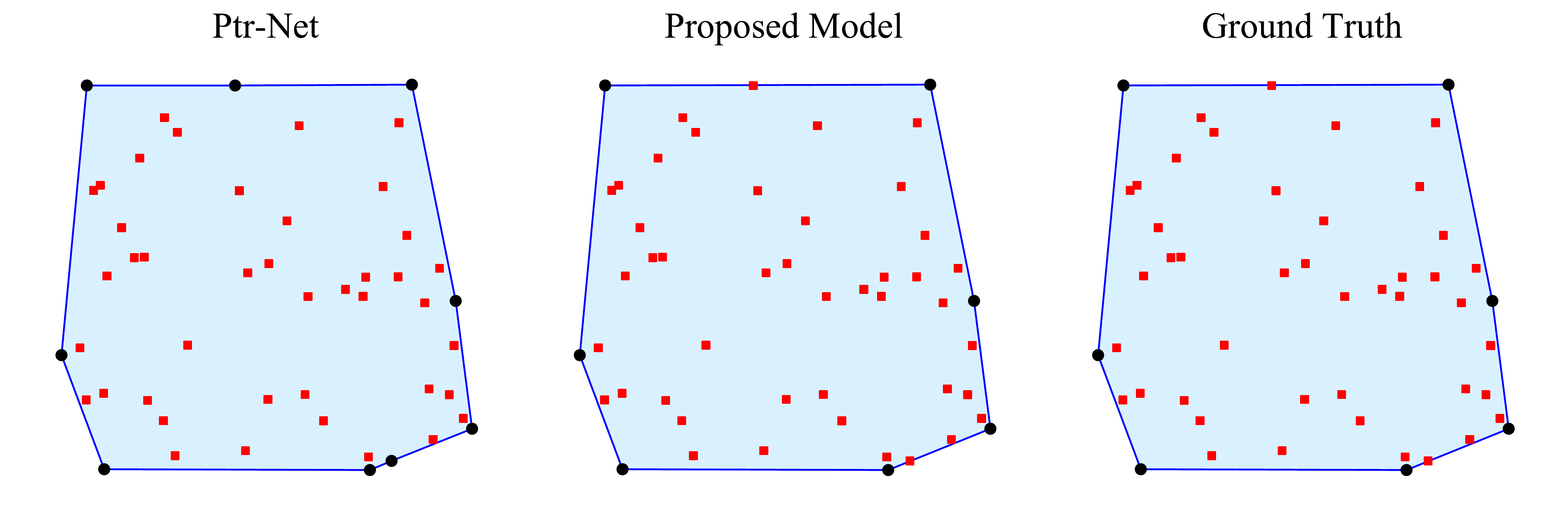}
\caption{Prediction example of the proposed model and the Ptr-Net compared to the ground truth for the convex hull problem when $m$=50. }
\label{fig:fig_88}
\end{figure}

{\noindent \bf  {Convex Hull.} }
Figure~\ref{fig:fig_conv_2} depicts the results of the proposed model compared with Ptr-Net when $m$=50 and 100. We observed that the proposed model outperforms Ptr-Net. Specifically, when $m$=100, the accuracy and area coverage of the proposed model is 80.00\% and 99.99\%, while Ptr-Net achieves 77.11\% and 99.99\%, respectively. Both models achieve nearly 100.00\% area coverage. Figure~\ref{fig:fig_88} shows one example when the output polygon of the proposed model is identical to the ground truth, while Ptr-Net fails to do so. For this example, the polygon formed by the output sequence of Ptr-Net has points that do not belong to the actual convex hull, and therefore the area coverage is 99.93\%. \\

\begin{table*}[t]
\centering
\begin{tabular} { l || c | c || c | c|| c| c}
\multirow{2}{*}{Method}  & \multicolumn{2}{c||}{TSP5} & \multicolumn{2}{c||}{TSP10} & \multicolumn{2}{c}{TSP50} \\       
                                                    &  ATL    & VTR    &  ATL  & VTR    &  ATL  & VTR    \\    
\hline \hline
Optimal & 2.12 & 100\% & 2.87  & 100\%  &  5.69 & 100\% \\
\hline 
A3 & 2.12 & 100\% & 2.87 & 100\%  &  5.81 & 100\% \\
OR tools & 2.13 & 100\% & 2.87 & 100\%  &  5.85 & 100\% \\
\hline 
Ptr-Net (SL) & 2.12 & 100\% & 2.88 & 96.5\%  &  5.88 & 68.2\% \\
GNN (SL) & 2.13 & 97.9\% & 2.89 & 62.6\%  &  6.57  & 0\% \\
GCN  (SL)& 2.12 & 100\% & 2.88 & 100\%  &  6.86 & 100\% \\
GAN (RL) & 2.13  & 100\%  & 2.87 & 100\%   &  5.83  & 100\%  \\
\textbf{Proposed  (SL)} & \textbf{2.12} & 100\% & \textbf{2.88} & 100\%  & \textbf{5.97} & 100\% \\
\end{tabular}
\caption{Average tour length (ATL) and valid tour rate (VTR) of the proposed model (greedy) compared to existing methods and models: A3 algorithm~\cite{2}, OR tools~\cite{40}, Prt-Net~\cite{2}, GNN~\cite{41}, GCN~\cite{14}, and GAN~\cite{13}. Here, SL and RL stand for supervised learning and reinforcement learning, respectively. } 
\label{tab:tab_2}
\end{table*}

{\noindent \bf  {TSP.} }
Table~\ref{tab:tab_2} presents the ATL and VTR of the proposed model (greedy) compared to existing methods (models).  For our experiments, the same training samples were used for a fair comparison. The proposed model outperforms other comparable models in terms of solution quality. The proposed model shows 100\% VTR in all cases, but the VTR of Ptr-Net and GNN-based model~\cite{41} were not 100\% VTR in most cases. For TSP50, the VTR of Ptr-Net and GNN-based model were only 68.2\% and 0\% respectively. Therefore, in the case of TSP50, even if the ATL of Ptr-Net is a little shorter than the proposed model, the VTR is too low for practical use. The proposed model showed the best performance among compared models for small-size TSP (TSP5), while the GAN-based model showed the best performance for TSP50.

In the case of the GNN-based model, the model produces the probability of going to the next city for each city.  It was observed that many invalid tours frequently occur because the next city is selected without considering previously visited cities. We also found that the results of the GCN-based model were not as good as those reported by~\cite{14}, especially for TSP50. We suspect that the quality of training samples may affect the performance.

\begin{figure*}[t]
\centering   
\vbox
{ 
\hbox
{
\subfigure[m=5] {
\includegraphics[width=5.8cm]{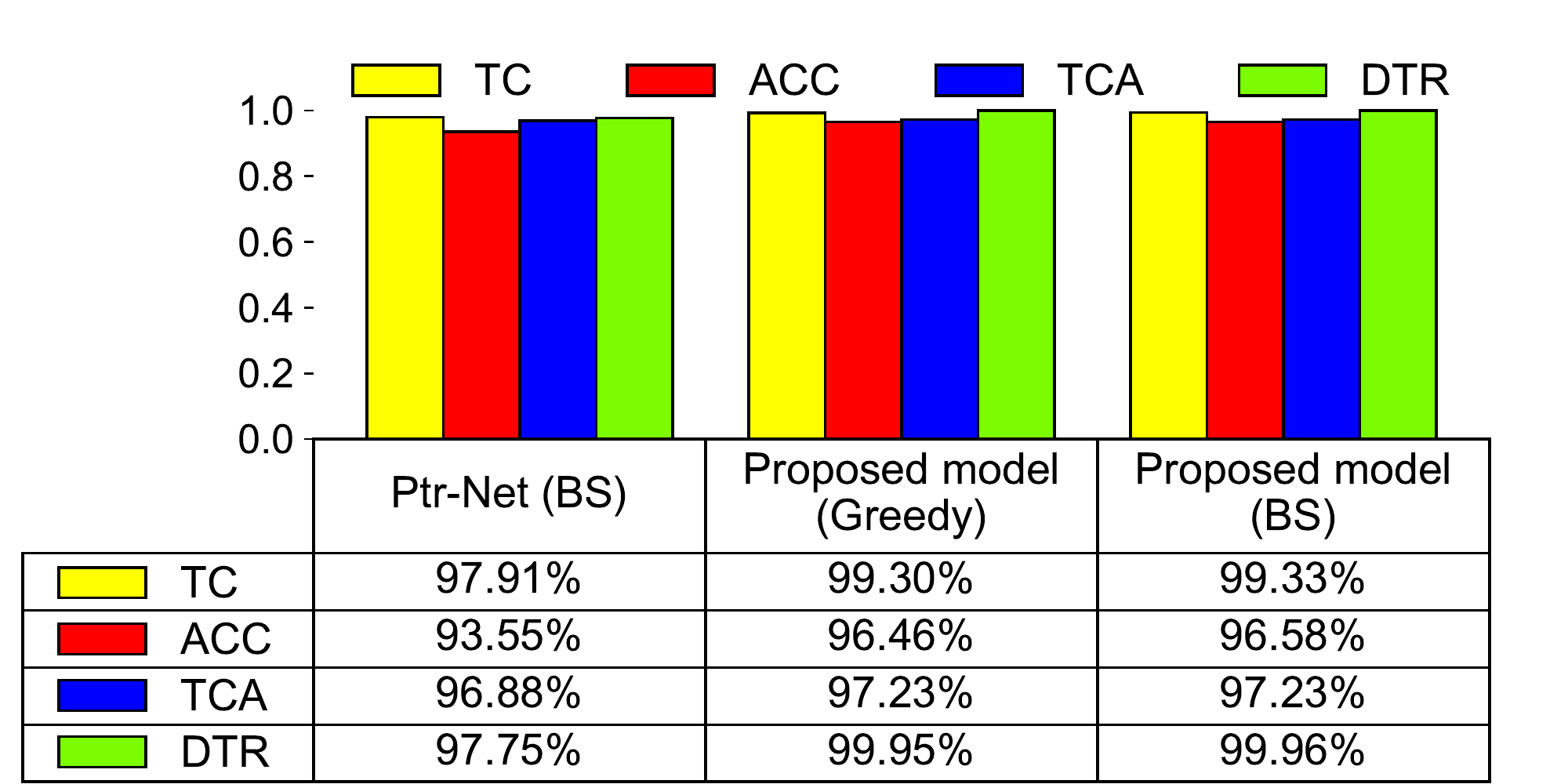}
}
\subfigure[m=10] {
\includegraphics[width=5.8cm]{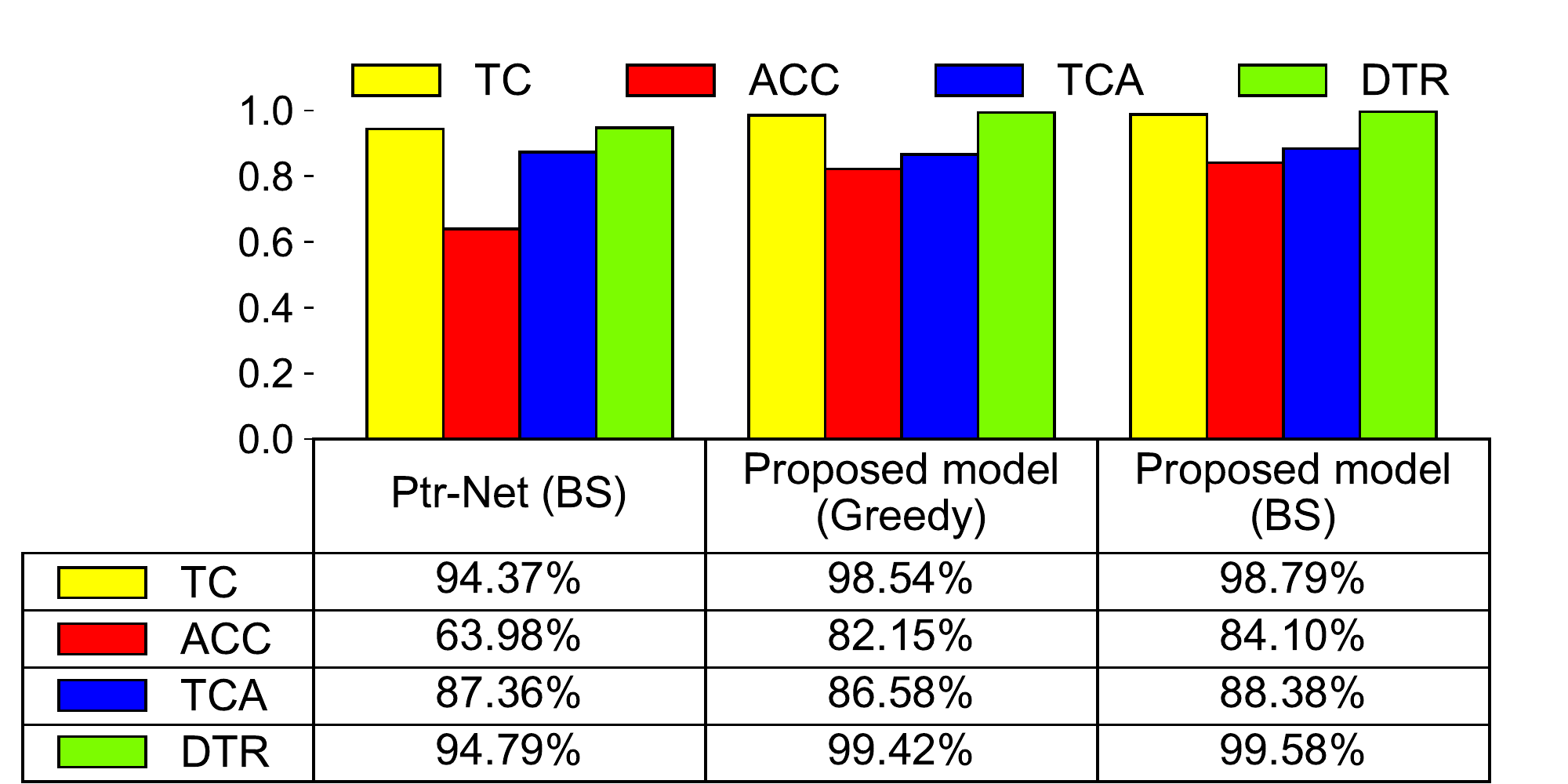}
}
}
\hbox
{
\subfigure[m=15]{
\includegraphics[width=5.8cm]{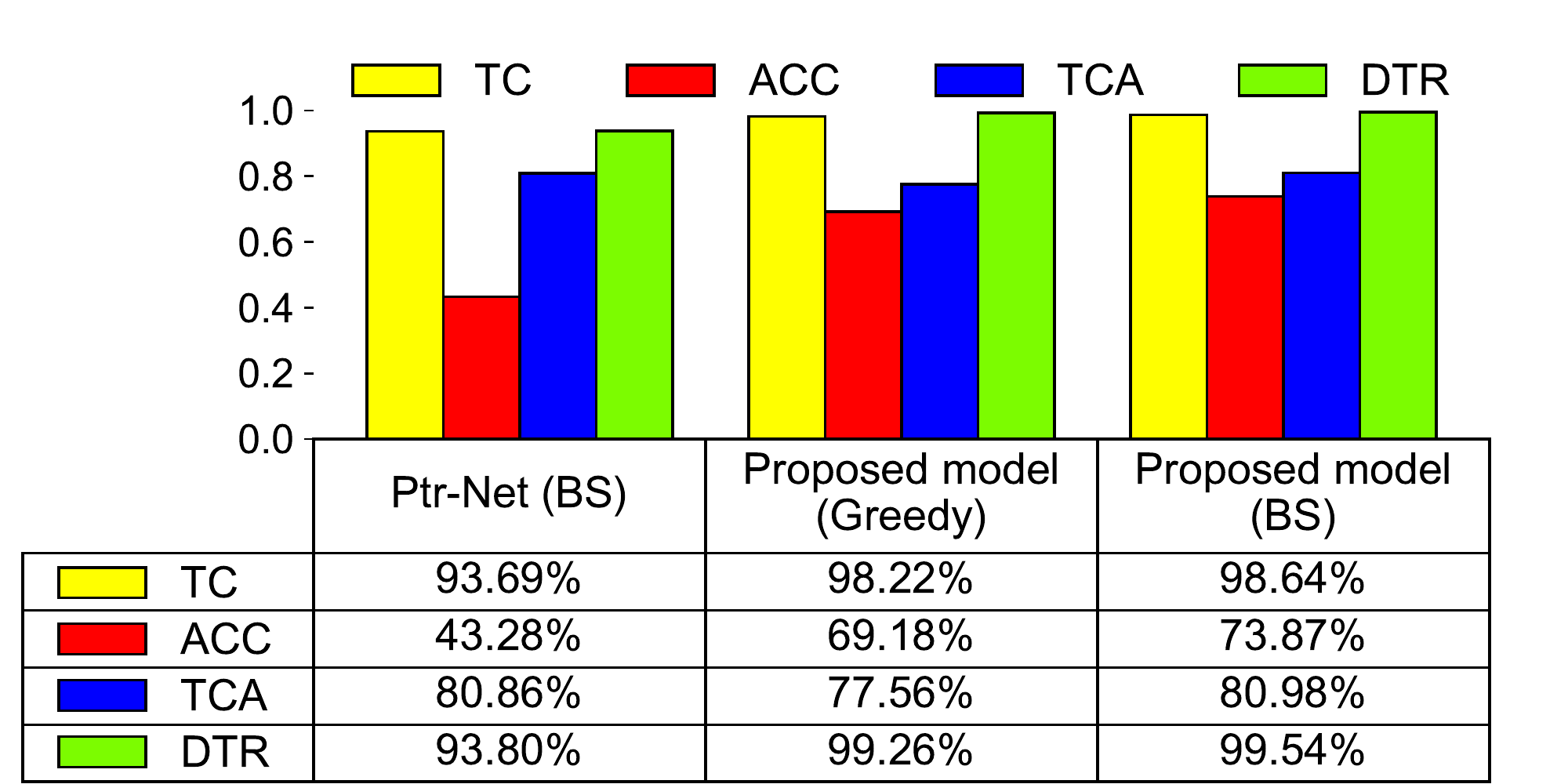}
}
\subfigure[m=20]{
\includegraphics[width=5.8cm]{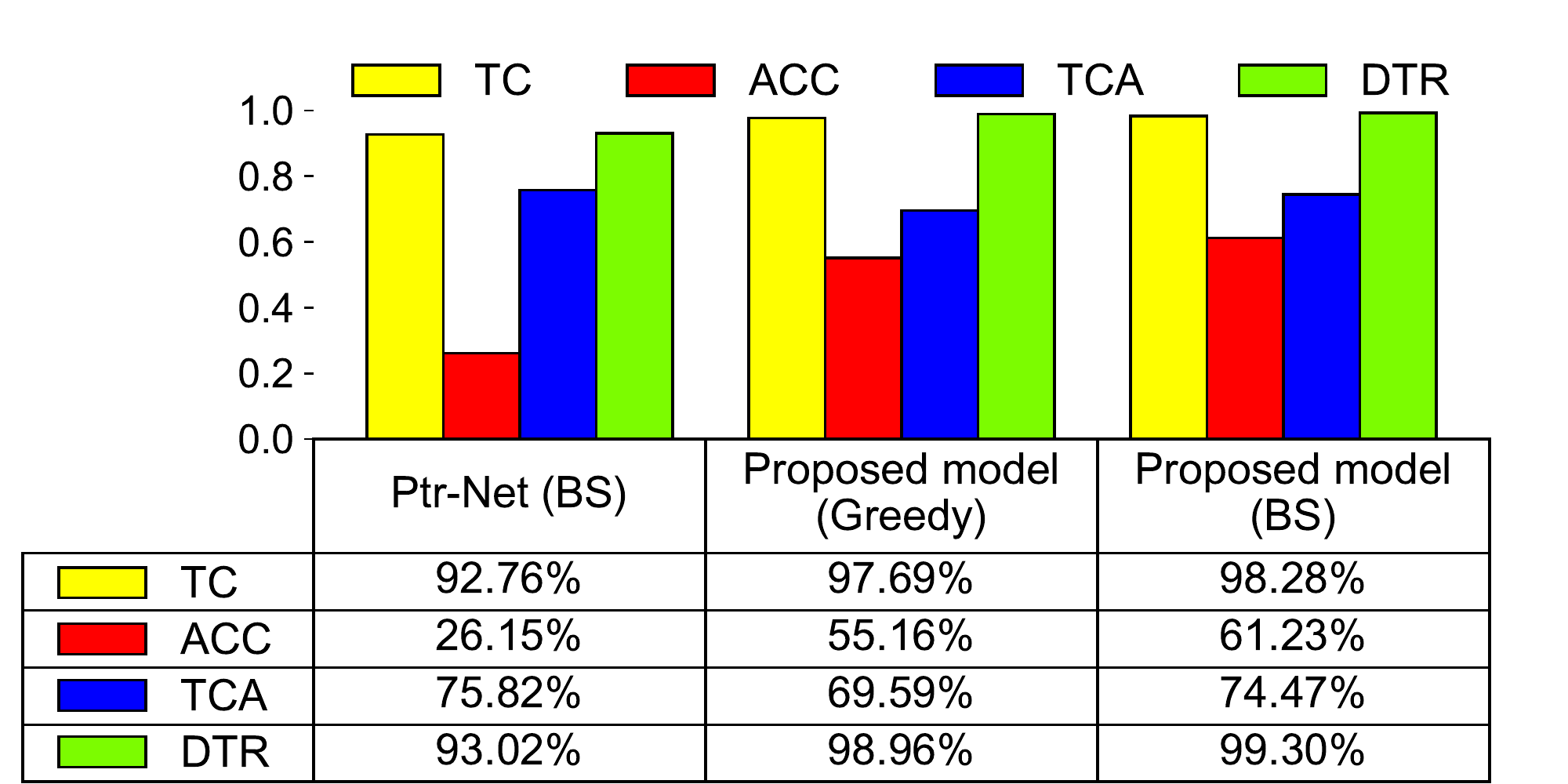}
}
}
}
\caption{Performance of the proposed model with greedy decoder and BS decoder for the DT problem.} 
  \label{fig:fig_9}
\end{figure*}

\subsection{Experiment 3: Effect of Beam Search Decoder}
\begin{figure}[t]
\centering
\includegraphics[width=12cm]{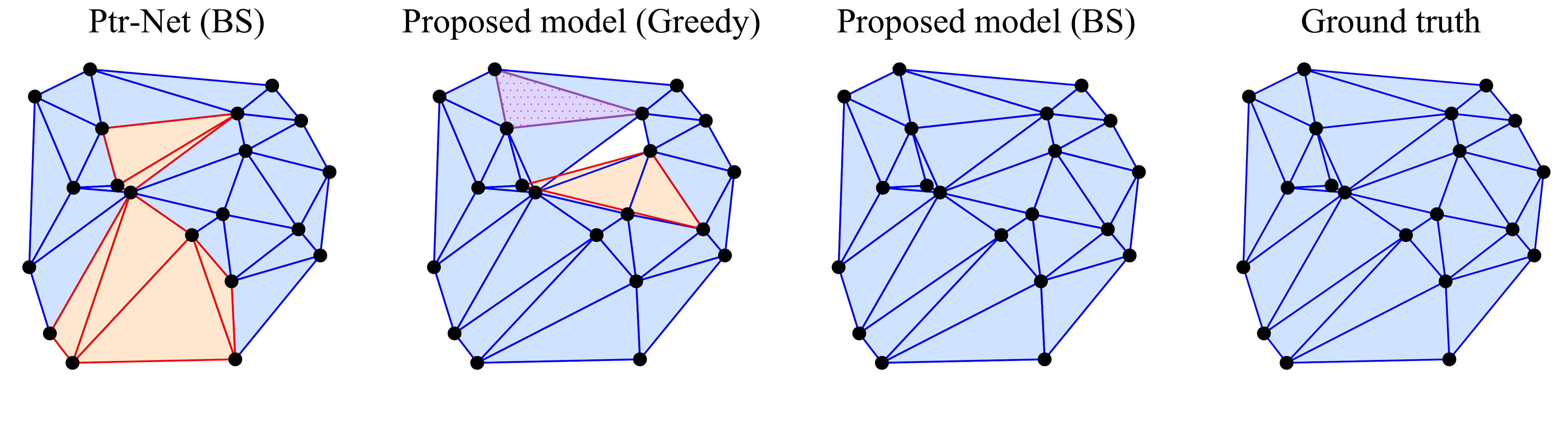}
\caption{Example of predictions of each model using the greedy and beam search (BS)  decoders when $m$ = 20. Blue elements represent elements in the ground truth, and red elements represent elements not in the ground truth. The dotted element is a replicated element. Only the proposed model (BS) predicts the ground truth.}
\label{fig:fig_10}
\end{figure}

{\noindent \bf  {DT.} }
Figure~\ref{fig:fig_9} shows the proposed model results when the BS decoder (BS = 4) is used instead of the greedy decoder. In the previous two experiments, the greedy decoder was used. We observe that the BS decoder significantly improves the performance compared to the greedy decoder in all four metrics. In particular, the effect of using the BS decoder increases as the $m$ value increases. For example, the performance gap between the greedy and BS decoder is only 0.03\% for the TC metric when $m$ = 5, and it increases to 0.59\% when $m$ =  20. For the ACC metric, the performance gap is 0.12\% when $m$ = 5, but it becomes 6.07\% when $m$ = 20. If the BS decoder is used, the performance can be improved by continuously tracking the most probable candidate points at the expense of prediction time. If the BS decoder is used together with the proposed masking scheme, the number of candidates of the first two steps (points) increases. Therefore, using the proposed masking scheme is more effective  because it offers more chances to determine the correct triangulation in the decoder. Figure~\ref{fig:fig_10} presents one example of predictions of each model with the BS decoder and the ground truth when $m$ = 20. Only the proposed model can predict the triangulation of the ground truth correctly. \\

\begin{figure}[t]
\centering
\hbox
{
\subfigure[m=50] {
\includegraphics[width=5.8cm]{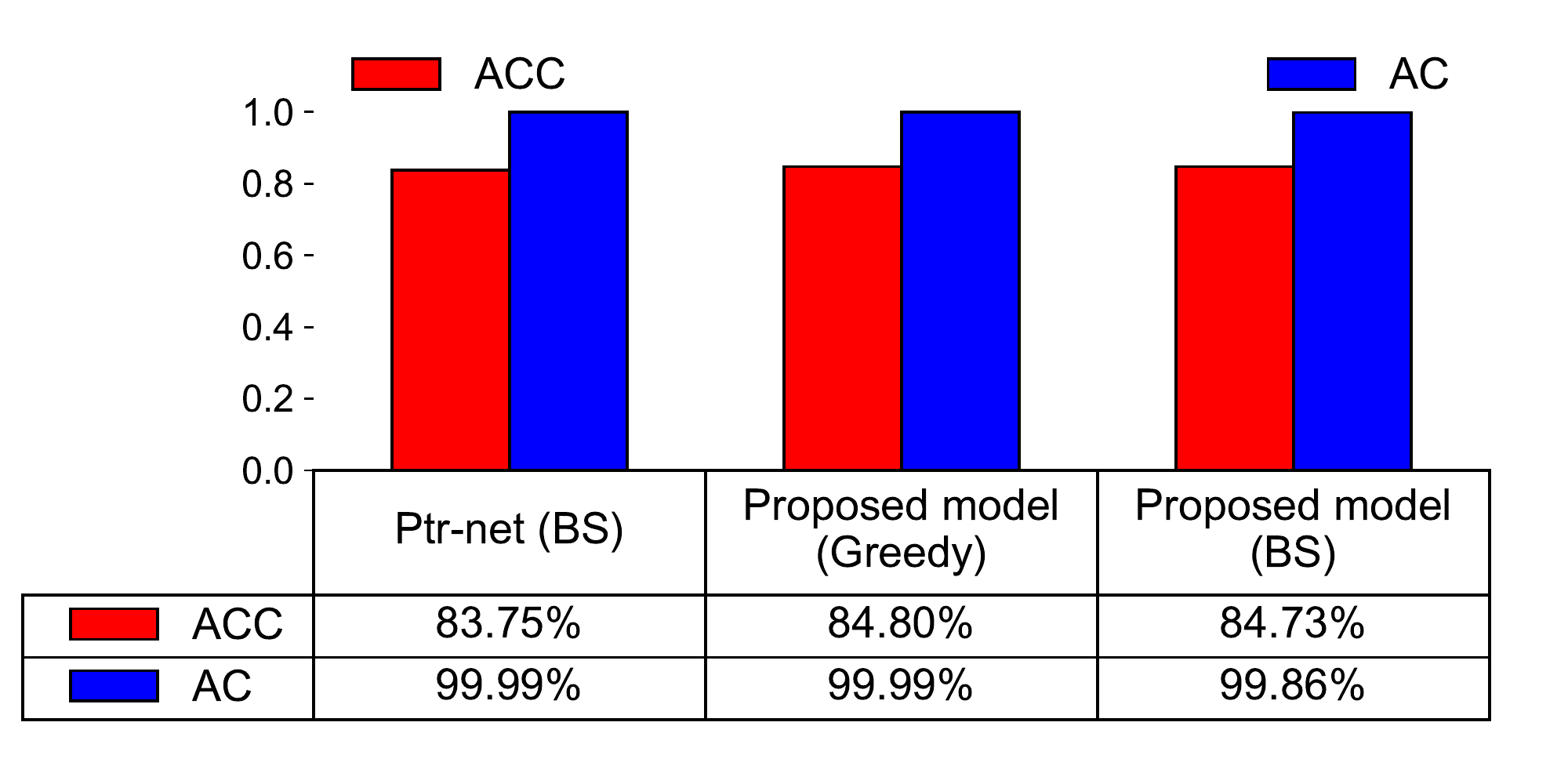}
}
\subfigure[m=100] 
{
\includegraphics[width=5.8cm]{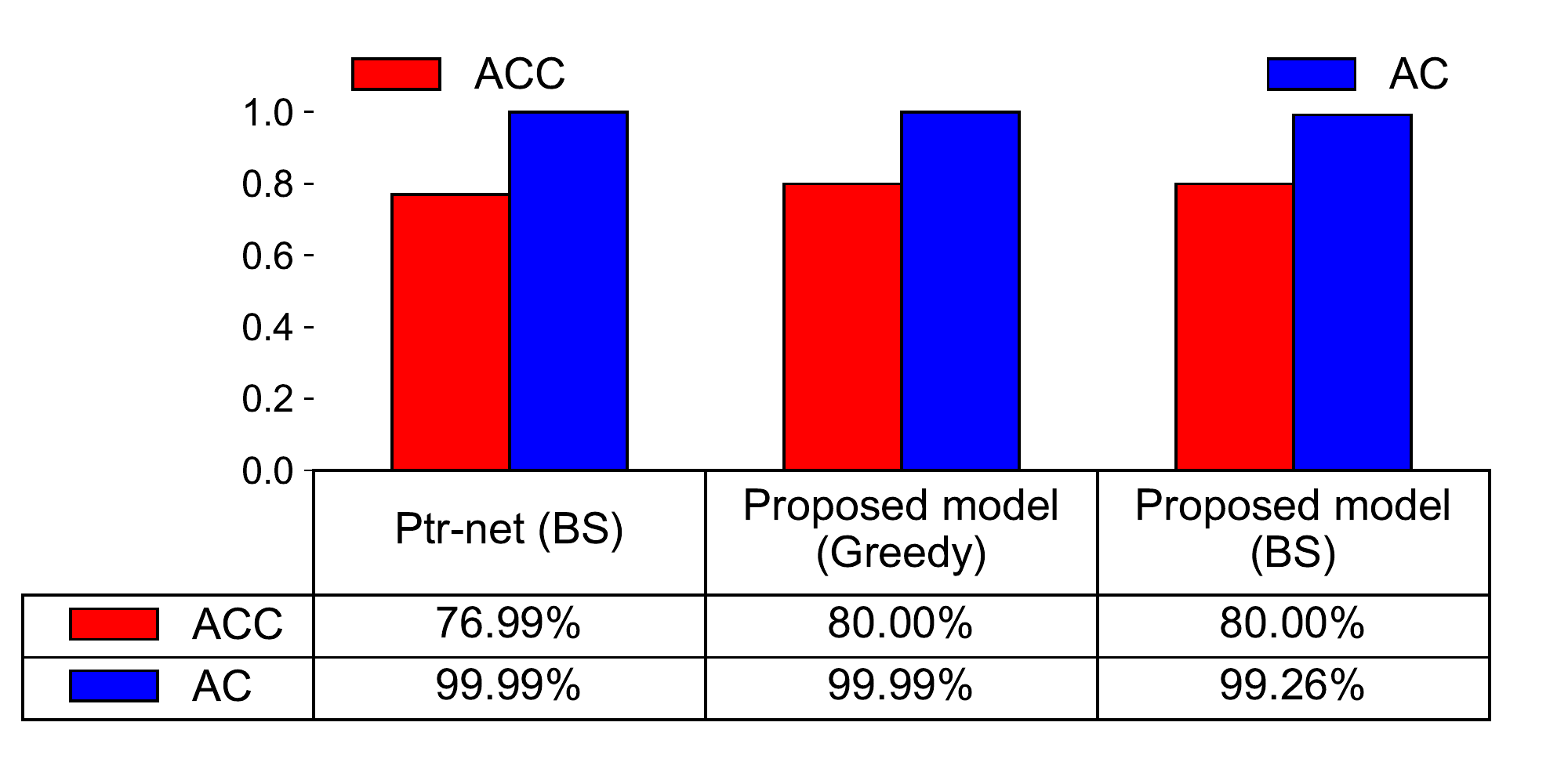}
}
}
\caption{Performance of the proposed model with greedy decoder and BS decoder for the convex hull problem.}
\label{fig:fig_conv_3}
\end{figure}

{\noindent \bf  {Convex Hull.} }
Figure~\ref{fig:fig_conv_3} shows the performance of the proposed model with greedy and BS decoders for the convex hull problem. Interestingly, unlike other examples, the performance slightly degrades with the BS decoder when $m$=50 and $m$=100. The model tends to prefer shorter output sequences when the BS decoders are used. We hypothesize that the BS decoder shows relatively poor performance compared with the greedy decoder because the model tends to select an output sequence that is shorter than the actual convex hull (ground truth). \\

\begin{figure}[t]
\centering
\includegraphics[width=10cm]{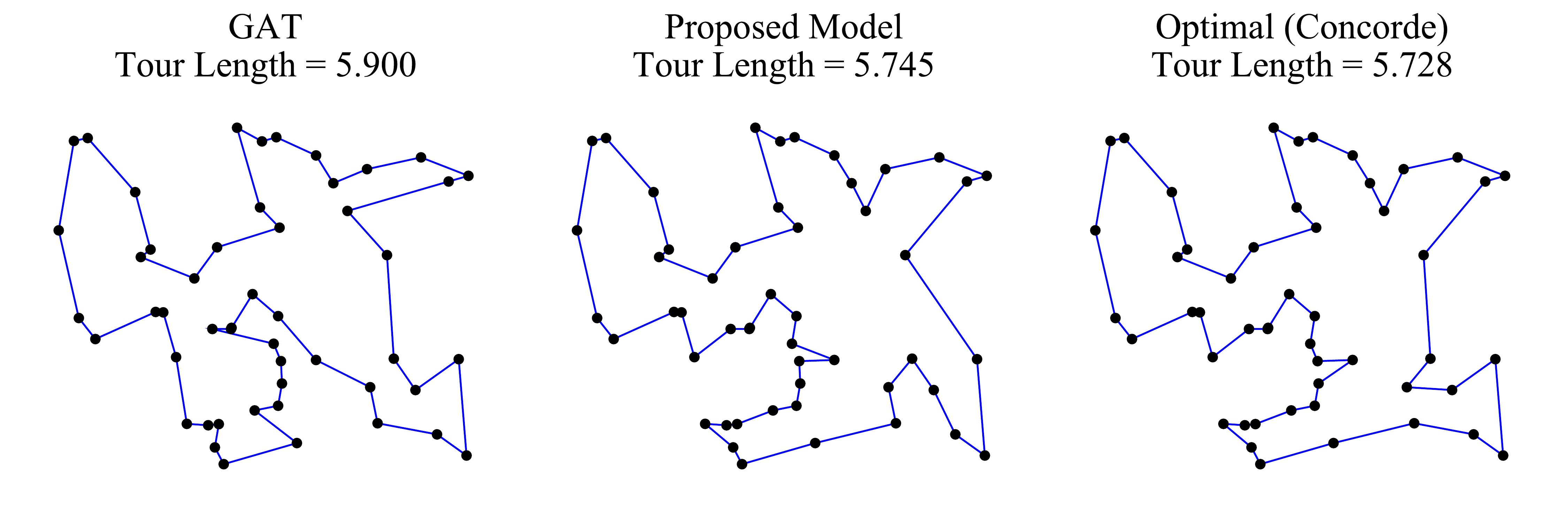}
\caption{Prediction example of the proposed model and the GAN-based model (\cite{13}) compared to the optimal tour (Concorde) for TSP50.}
\label{fig:fig_21}
\end{figure}

{\noindent \bf  {TSP.} }
Table~\ref{tab:tab_3} presents the ATL and VTR of the proposed model (BS) compared with existing methods (models). For a fair comparison, all models use the same beam width, i.e., four. First, the proposed model with the BS decoder showed the best performance among supervised learning-based models. For TSP50, the proposed model with the BS decoder that selects the shortest tour even outperforms the algorithm-based methods, i.e., A3 algorithm and OR tools. Second, the proposed model is based on the supervised learning, but it shows comparable results with the reinforcement learning-based model~\cite{13}. Third, the beam search strategy that selects the shortest tour among the final beam candidate sequences boosts our performance. This beam search strategy is more effective for large-size TSP (TSP50). Finally, we observed that the proposed model with the BS decoder showed better performance than that with the greedy decoder in all cases.

Figure~\ref{fig:fig_21} shows an example wherein the proposed model shows better output performance compared with the GAN-based model for TSP50. The ATL of the proposed model is 5.745, but the ATL of the GAN-based model is 5.900. For this example, the optimal tour length of Concorde is 5.728.

\begin{table*}[t]
\centering
\begin{tabular} { l || c | c || c | c|| c| c}
\multirow{2}{*}{Method}  & \multicolumn{2}{c||}{TSP5} & \multicolumn{2}{c||}{TSP10} & \multicolumn{2}{c}{TSP50} \\       
                                                    &  ATL    & VTR    &  ATL  & VTR    &  ATL.  & VTR    \\    
\hline 
Optimal & 2.12 & 100\% & 2.87 & 100\%  &  5.69  & 100\% \\
\hline 
A3  & 2.12 & 100\% & 2.87 & 100\%  &  5.81 & 100\% \\
OR tools & 2.13 & 100\% & 2.87 & 100\%  &  5.85& 100\% \\
\hline 
Ptr-Net (SL, BS) & 2.12 & 100\% & 2.88 & 97.0\%  &  5.81 & 84.6\% \\
GNN (SL, BS) & 2.17 & 93.2\% & 2.89 & 71.2\%  &  6.60  & 0\% \\
GCN (SL, BS) & 2.12 & 100\% & 2.87 & 100\%  &  6.32  & 100\% \\
GAN (RL, S) &  2.12  &   100\% & 2.87    &   100\%     &     5.80      &  100\% \\
\textbf{Proposed (SL, BS)} & \textbf{2.12} & 100\% & \textbf{2.87} & 100\%  &  \textbf{5.85} & 100\% \\
\textbf{Proposed (SL, BSS)} & \textbf{2.12} &  100\% &  \textbf{2.87} & 100\%  &  \textbf{5.80} & 100\% \\
\end{tabular}
\caption{Average tour length (ATL) and valid tour rate (VTR) of the proposed model (BS) compared to existing methods and models: A3 algorithm~\cite{2}, OR tools~\cite{40}, Prt-Net~\cite{2}, GNN~\cite{41}, GCN~\cite{14}, and GAN~\cite{13}. In the table, SL: supervised learning, RL: reinforcement learning, S: Sampling, BS: beam search, and BSS: beam search with the shortest tour.}
\label{tab:tab_3}
\end{table*}

\begin{figure}[ht]
\centering
\hbox
{
\subfigure[The first edge predicted by the model is not included in the ground truth] {
\includegraphics[width=5.5cm]{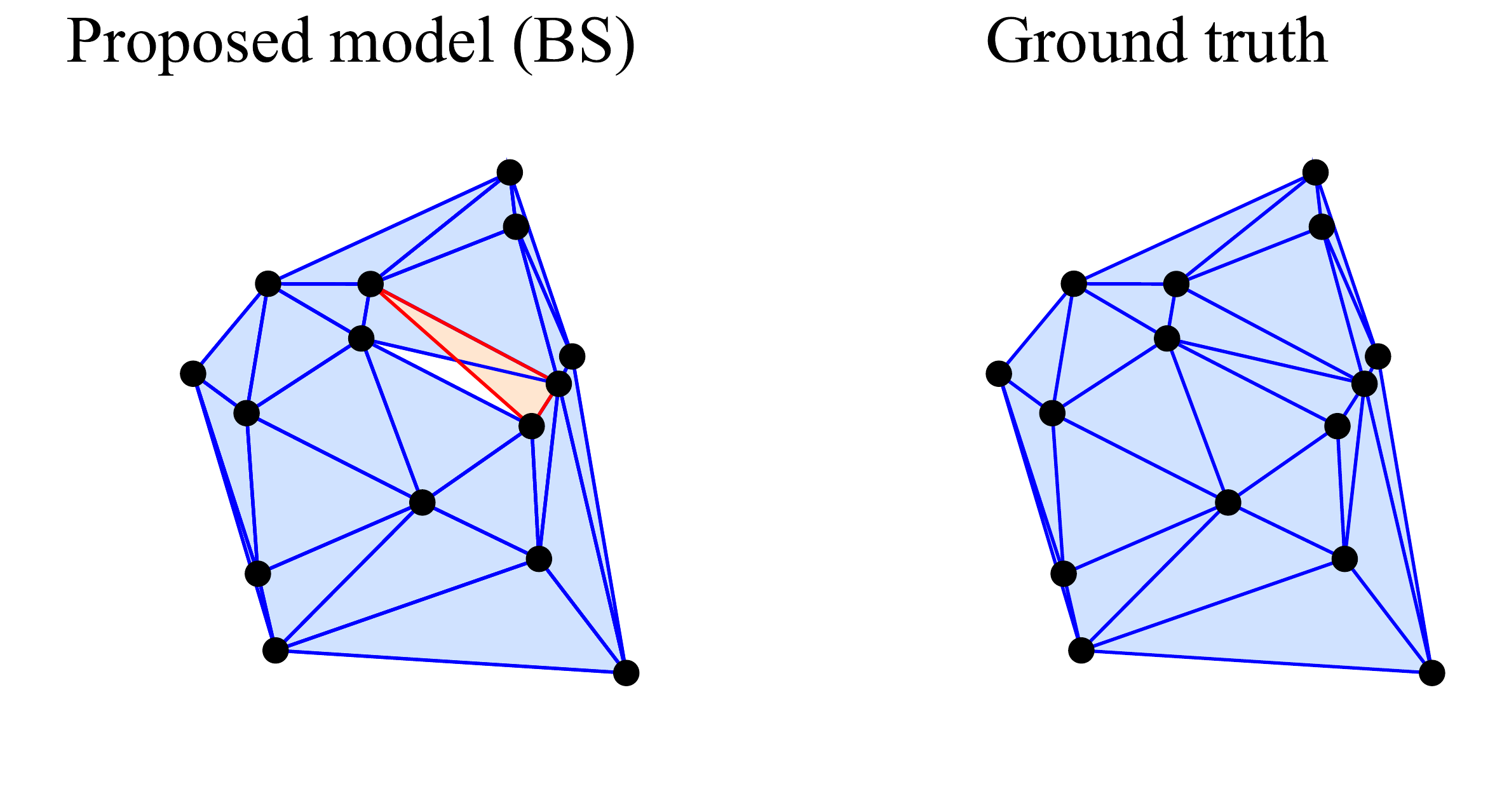}
}
\subfigure[The duplicated triangle (dotted) is predicted by the model] 
{
\includegraphics[width=5.5cm]{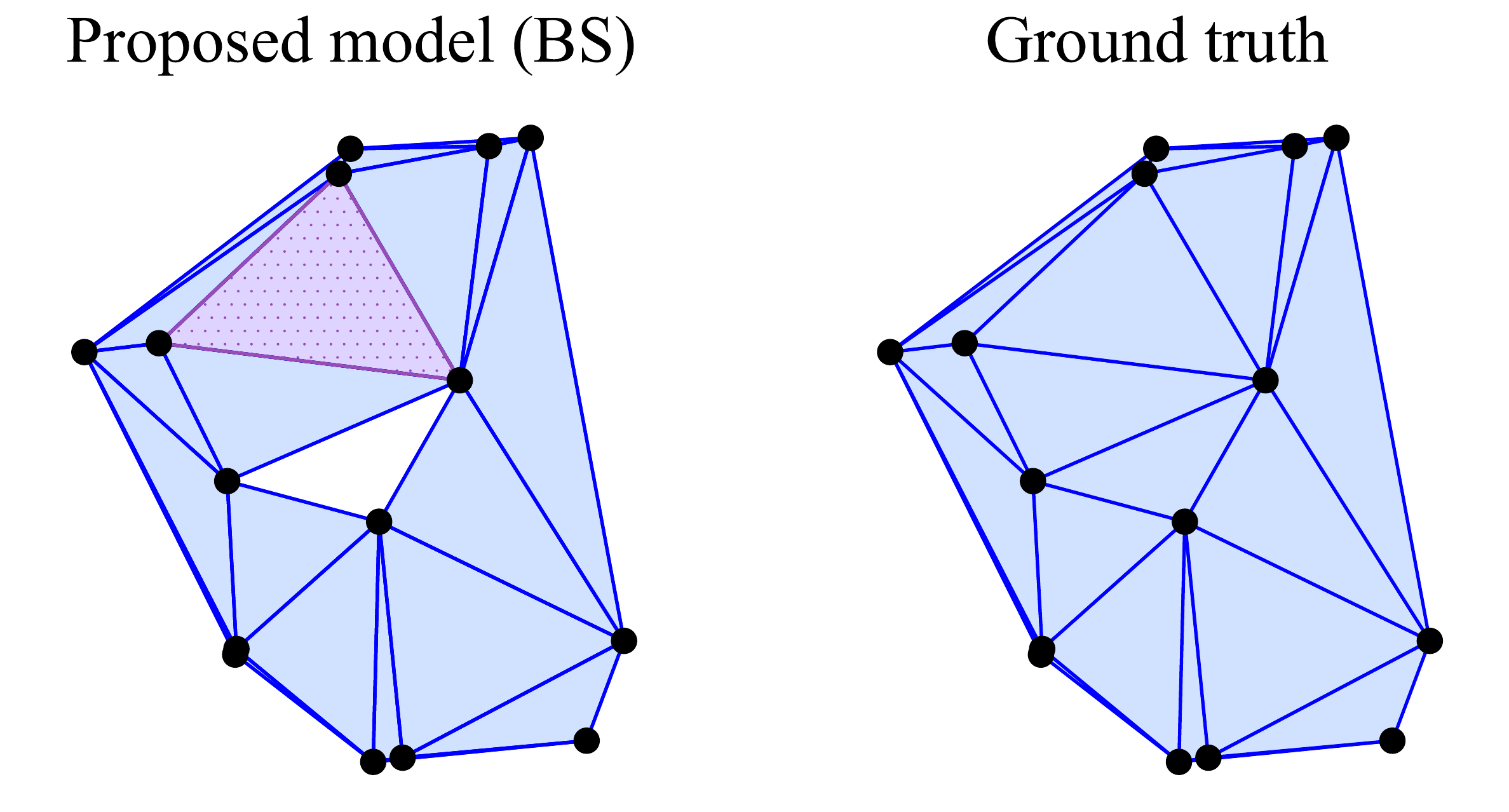}
}
}
\caption{Two examples of when the proposed model fails for the DT problem.}
\label{fig:fig_12}
\end{figure}

\begin{figure}[t]
\centering
\includegraphics[width=7cm]{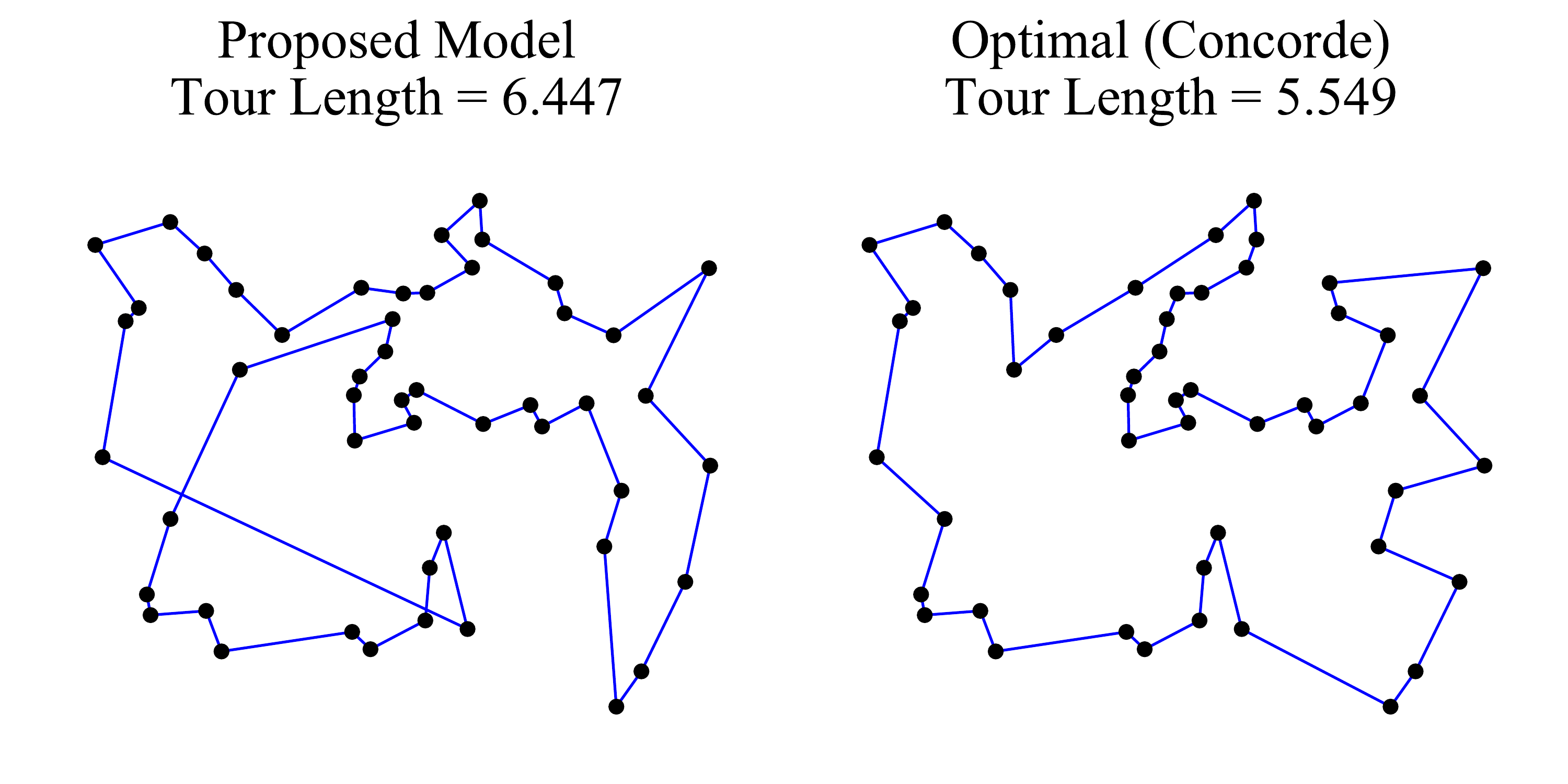}
\caption{One example when the proposed model prediction shows poor performance for TSP50. }
\label{fig:fig_55}
\end{figure}

\section{Discussion}
\label{sec:Discussion}
We investigate scenarios wherein the model's prediction shows poor performance. We first present several cases of failure of the predictions of the proposed model for the DT problem. First, the edge consisting of the first two predicted points is an edge that is not in the ground truth. For this case, the proposed masking scheme cannot exclude these points from the candidates because it evaluates the candidate points every three steps. We observe that these failure cases often occur when the testing sample points have similar or identical values in the $x$-axis. For these cases, the model learning performance can deteriorate because the input sequence is lexicographically ordered. The model also fails to predict the ground truth when it predicts the same triangles multiple times, especially for large m values. In addition, the number of triangles predicted by the model is sometimes more or less than the number of triangles in the ground truth. Figure~\ref{fig:fig_12} depicts two cases in which the model prediction fails. In the first case, the first edge (red) predicted by the model is not in the ground truth, as described earlier. It creates an initial triangle that is not in the ground truth. In the second case, the model predicts the same triangles (hatched element) twice, but not the triangle in the ground truth (white element). Future work will also study how to solve these cases in the decoder.

For TSP, we observed that the proposed model does not produce an invalid tour, but it sometimes produces relatively long tours compared with the optimal tour. Figure~\ref{fig:fig_55} shows one such example. The tour length of the proposed model is 6.447, while the tour length of the Concorde is 5.549. When the tour length predicted by the model is relatively long as in this example, an intersection between the edges between distant cities occurs, and the tour length increases.

Table~\ref{tab:tab_4} presents the training time comparison with existing methods (models) for 10,000 samples. We use a serial implementation of the Concorde and A3 algorithms for the TSP solver. The GCN-based model has the lowest convergence speed with long training time compared with other models. The training speed of the proposed model is approximately 70 times higher than that of the GCN-based model.

In terms of solution (inference) time, exact solutions exist for convex hull and DT problems, while the proposed model produces approximate solutions. Therefore, solution times for these problems are not compared. For TSP, finding exact solutions is very costly for large-size problems (e.g., TSP50). For such problems, we compare the solution time of the deep learning model and Concorde~\cite{42}, which is considered to produce the fastest optimal TSP solution. Similar to~\cite{14}, we use the total wall clock time taken to solve 10,000 test samples. Note that the solution time can vary depending on which deep learning library (e.g., PyTorch or TensorFlow) is used as well as the implementation language (e.g., C or Python).

Table~\ref{tab:tab_5} lists the solution time comparison with existing methods for TSP50. A beam width (or sampling size) of four is used for the beam search decoder. We observed that the proposed model achieves a faster solution time compared with traditional algorithm-based solvers like Concorde and A3. Among compared models, the GCN-based model was the slowest.

\begin{table}[t]
\centering
\begin{tabular} { l ||  c| c| c | c | c}
Method & Ptr-Net & GNN & GCN & GAN &\textbf{Proposed} \\    
\hline   
Time & 2.6s & 3.5s & 191.0s & 4.3s & \textbf{2.7s} \\
\end{tabular}
\caption{Training time for 10,000 TSP samples: Prt-Net~\cite{2}, GNN~\cite{41},  GCN~\cite{14}, and GAN~\cite{13}.} 
\label{tab:tab_4}
\end{table}

\begin{table}[t]
\centering
\begin{tabular} { l || c | c || c| c| c | c| c}
Method & Concorde & A3 & Ptr-Net & GNN & GCN & GAN & \textbf{Proposed} \\    
\hline   
Time  & \multirow{2}{*}{531.1s} & \multirow{2}{*}{ 51.7s} & 19.1s & 24.9s & 60.0s & 1.0s& \textbf{21.9s}\\
Time (BS) &  & & 27.1s & 25.3s& 65.0s & 2.0s & \textbf{30.2s} \\
\end{tabular}
\caption{Inference time to solve 10,000 test samples for TSP50: Concorde~\cite{42}, A3 algorithm~\cite{2}, Prt-Net~\cite{2}, GNN~\cite{41}, GCN~\cite{14}, and GAN~\cite{13}. The result in the first row is the result using the greedy decoder. } 
\label{tab:tab_5}
\end{table}

\section{Conclusions}
\label{sec:Conclusion}
We propose a novel neural network model for learning COPs involving geometry using new attention mechanisms based on self-attention and domain knowledge. We show that the proposed neural net can be used to learn and produce competitive approximate solutions for various COPs involving geometry. In particular, our experiments show that the proposed model learns and produces a competitive TSP approximate solver. For TSP, the proposed model outperforms other recent state-of-the-art supervised learning-based models in terms of solution quality and speed.

Our experiments show that the effect of employing self-attention in the encoder is effective for the model to learn point-to-point relationships and is most effective for the DT problem among three problems. For the DT problem, it is important for the model to learn about the relationship among points. Employing self-attention in the encoder helps to better learn topological and geometric relationships among points. Second, we observe that the input/output sequence order matters for all three problems; the model learns much faster and better using the proposed input/output sequence ordering. In particular, we observe that the sequence ordering is more important as the sequence length increases. Finally, we observed that the proposed masking scheme is effective for improving the performance of the model as it excludes candidates that do not satisfy the geometric requirements. For the convex hull problem, we simply check whether the candidate points satisfy the convexity condition or appeared previously in the output sequence. Therefore, the polygon formed by the output sequence could not be a convex hull although it satisfies the convexity requirement. We expect better output performance if other masking schemes are used for the convex hull problem. 

Future work will apply the proposed approach to other well-known COPs involving geometry such as the vehicle routing problem. We also plan to apply the multi-head attention technique used in the transformer model to our model to further improve its performance.

\section*{CRediT authorship contribution statement}
\textbf{Jaeseung Lee:} Software, Methodology, Investigation, \textbf{Woojin Choi:} Software, Investigation, \textbf{Jibum Kim:} Writing - Original Draft, Writing - Reviewing and Editing, Supervison
\section*{Declaration of Competing Interest}
The authors declare that they have no known competing financial interests or personal relationships that could have appeared to influence the work reported in this paper. 

\section*{Acknowledgments}
This work was supported in part by the National Research Foundation of Korea (NRF) Grant funded by the Korea Government (MSIT) under Grant NRF-2020R1A2C1\\007917. This work was also supported in part by the  National Research Foundation of Korea (NRF) Grant funded by the Korea Government (MSIT) under Grant NRF-2022R1A4A5034121.

\bibliography{library}


\end{document}